\documentclass[twocolumn]{aastex62}\submitjournal{ApJ}\received{2018 April 27}\revised{2018 July 27}\accepted{2018 August 20}

\newcommand{\secref}[1]{\S\ref{#1}}
\newcommand{\colspace}{3.0pt}
\newcommand{\cpr}{0.12^2}
\newcommand{\cpa}{0.1^2}
\newcommand{\csr}{0.095^2}
\newcommand{\csa}{0.08^2}
\newcommand{\cdd}{0.11^2}

\newcommand{\cdf}{0.025^2} 
\newcommand{\ccf}{0.02^2}
\newcommand{\logt}{\log_{10}}
\newcommand{\htwo}{\rm H_2}
\newcommand{\rtf}{r$_{25}$}
\newcommand{\metal}{\rm 12+\logt({\rm O/H})}

\newcommand{\CASS}{\affiliation{Center for Astrophysics and Space Sciences, Department of Physics, University of California, San Diego\\9500 Gilman Drive, La Jolla, CA 92093, USA}}
\newcommand{\OSU}{\affiliation{Department of Astronomy, The Ohio State University\\4055 McPherson Laboratory, 140 West 18th Ave, Columbus, OH 43210, USA}}
\newcommand{\CIERA}{\affiliation{CIERA and Department of Physics and Astronomy, Northwestern University\\2145 Sheridan Road, Evanston, IL 60208, USA}}

\shorttitle{The Spatially Resolved Dust-to-metals Ratio in M101}
\shortauthors{Chiang et al.}

\begin{document}
\title{The Spatially Resolved Dust-to-metals Ratio in M101}

\correspondingauthor{I-Da Chiang}
\email{idchiang@ucsd.edu}

% Note: CJK package is not working after AASTEX v6.1. Can ask CJK names but don't expect it to work
%
\author[0000-0003-2551-7148]{I-Da Chiang}
\CASS
\author[0000-0002-4378-8534]{Karin M. Sandstrom}
\CASS
\author[0000-0002-5235-5589]{J\'{e}r\'{e}my Chastenet}
\CASS
\author[0000-0001-6421-0953]{L. Clifton Johnson}
\CIERA
\author[0000-0002-2545-1700]{Adam K. Leroy}
\OSU
\author[0000-0003-4161-2639]{Dyas Utomo}
\OSU

% Note: Try not to have any references in abstract. Didn't find this rule in AAS guide, but no harm to avoid this.
% Note: "(acronyms) Terms defined in the abstract should be defined independently in the main text". Actually, just no acronyms in abstract...
\begin{abstract}
The dust-to-metals ratio describes the fraction of the heavy elements contained in dust grains, and its variation provides key insights into the life cycle of dust. We measure the dust-to-metals ratio in M101, a nearby galaxy with a radial metallicity ($Z$) gradient spanning $\sim$1 dex. We fit the spectral energy distribution of dust from 100 to 500 \micron\ with five variants of the modified blackbody dust emission model in which we vary the temperature distribution and how emissivity depends on wavelength. Among them, the model with a single-temperature blackbody modified by a broken power-law emissivity gives the statistically best fit and physically most plausible results. Using these results, we show that the dust-to-gas ratio is proportional to $\rm Z^{1.7}$. This implies that the dust-to-metals ratio is not constant in M101, but decreases as a function of radius, which is equivalent to a lower fraction of metals trapped in dust at low metallicity (large radius). The dust-to-metals ratio in M101 remains at or above what would be predicted by the minimum depletion level of metals observed in the Milky Way. Our current knowledge of the metallicity-dependent CO-to-H$_2$ conversion factor suggests that variations in the conversion factor cannot be responsible for the trends in dust-to-metals ratio we observe. This change of dust-to-metals ratio is significantly correlated with the fraction of molecular hydrogen, which suggests that the accretion of gas-phase metals onto existing dust grains could contribute to a variable dust-to-metals ratio.
\end{abstract}

% galaxies: evolution --- galaxies: abundances 

\section{Introduction} \label{sec: intro}
% Why we should care about Dust
Interstellar dust grains participate in many important physical and chemical processes in the interstellar medium (ISM). For example, the surface of dust is the catalyst for formation of some molecules, especially $\htwo$ \citep{GOULD63, CAZAUX04}. Dust also shields gas from the interstellar radiation field (ISRF), and allows the low temperatures crucial to star formation to emerge deep within molecular clouds \citep{KRUMHOLZ11, YAMASAWA11, GLOVER12}. Dust plays an important role in the observed spectral energy distribution (SED) of galaxies: it absorbs and scatters starlight, and reemits the absorbed energy at infrared (IR) wavelengths \citep{CALZETTI01, BUAT12}. Thus, it is important to understand the properties of dust before we can fully understand the ISM and the observed SED from galaxies.

% Dust formation mechanism. Ends at "destroy of dust & dust formation from ISM" depend on metallicity
The amount of interstellar dust depends on the balance between dust formation and dust destruction. The mechanisms of dust destruction include supernovae (SNe) shocks, thermal evaporation, cosmic rays, and dust incorporated into newly formed stars \citep{DWEK98, HIRASHITA99}. The dust formation mechanisms include accretion of metals in the ISM onto existing dust grains, formation of new dust grains in the winds of AGB stars, and dust formation in type II SNe \citep{DWEK98, ASANO13}. Different dominant dust destruction and formation mechanisms would result in a different dust-to-gas mass ratio (DGR):
\begin{equation}
{\rm DGR} \equiv \Sigma_d / \Sigma_{\rm gas}
\end{equation}
and dust-to-metals ratio (DTM):
\begin{equation}
    {\rm DTM} \equiv {\rm DGR} / Z,
\end{equation}
where $\Sigma_d$ is the dust mass surface density, $\Sigma_{\rm gas}$ is the total gas mass surface density, which includes the contribution from \textsc{HI}, H$_2$ and He, and $Z$ is the metallicity. Note that some authors replace $\Sigma_{\rm gas}$ with hydrogen mass surface density in the definition of DGR, e.g., \citet{DRAINE14, GORDON14}. Other than the formation and destruction mechanisms affecting DGR and DTM, the DTM itself can directly impact the ISM dust accretion rate \citep{DWEK98}. Thus, studying DGR and DTM provides key insights into the dust life cycle.

% What it might mean by having a constant/varying/broken-power dust-to-metals relation.
Theoretical dust life cycle models yield varying predictions for the DTM as a function of metallicity and local environment. Models in \citet{SODROSKI97, DWEK98} show that the DGR gradient scales linearly with the metallicity gradient, and the DTM is nearly a constant. This can be achieved by a constant rate of dust formation and destruction, which results in a constant fraction of metal incorporated into dust, and thus DTM at all chemical evolution stages is a constant \citep{GALLIANO08}. Other studies show that DTM is not always a constant, but a multi-stage variable as metallicity increases. At low metallicity, ISM accretion is less effective and the dust production rate is dominated by stellar ejecta, which could result in a locally constant DTM in this low metallicity regime \citep{HIRASHITA11}. Above a certain critical metallicity, the efficiency of dust accretion may increase, which would result in a DTM increasing with metallicity \citep{ZHUKOVSKA08, HIRASHITA11, FELDMANN15}. The critical metallicity depends on model and choices of parameters, and usually falls in the range of $\metal = 7.5$ and $8.5$ \citep{HIRASHITA99, ZHUKOVSKA08, HIRASHITA11, ASANO13, ZHUKOVSKA16}.

% Observation supporting constant DTM.
Several observational studies support a constant DTM. In \citet{ISSA90}, the authors collated the DGR gradients and metallicity gradients from previous studies in M31, M33, and M51, and reached the conclusion that the slopes of DGR and metallicity with galactic radius are consistent with each other. In \citet{LEROY11}, the authors followed the approaches in \citet{DRAINE07} to derive the dust masses in local group galaxies. They showed that DTM is a constant across $8.0 \la \metal \la 9.0$. In \citet{DRAINE14}, the authors fit the IR SED in M31 to a renormalized version of dust model described in \citet{DRAINE07}. The authors showed that their derived DGR scales linearly with metallicity where metallicity measurements are reported by \citet{ZURITA12}. Importantly, the relation between dust and metallicity is consistent with $M_d/M_H \sim 0.0091 Z/Z_\sun$, a prediction from depletion conditions in the cloud toward $\zeta$Oph in the Milky Way (MW) \citep{DRAINE11, DRAINE14}.

% Observation supporting variable DTM.
There are also observational results supporting a varying DTM. In \citet{LISENFELD98}, the authors studied the DTM in 44 dwarf galaxies, and found a varying DTM. In \citet{HIRASHITA02}, the authors study 16 blue compact dwarf (BCD) galaxies, and found that $\logt(\rm DGR)$ spreads from $-3.3$ to $-4.6$ within $\rm 7.9 < \metal < 8.6$, indicating an variable DTM because the slope between DGR and metallicity is not unity. The authors hypothesized that this phenomenon is the result of the variation in dust destruction efficiency by SNe, which depends on the star formation history of the region. \citet{HUNT05} also showed a 2 dex spread of DGR at $8 \leq \metal \leq 9$. They also reported that the BCD SBS 0335$-$052, which has a metallicity $\metal = 7.32$, has an extremely low dust mass, two orders of magnitude below a linear trend with metallicity. Similarly, \citet{HERRERA-CAMUS12} and \citet{FISHER14} showed that the local dwarf galaxy I Zw 18 has a DGR two orders of magnitude below the linear trend derived from local galaxies. In \citet{REMY-RUYER14}, the authors compiled DGR measurements for 126 galaxies, with 30\% of their sample having $\metal\leq 8.0$. They showed that there might be a discontinuity of the linear DTM at oxygen abundance $\metal = 8$, and the galaxies below that metallicity have $\rm DGR \propto Z^{3.1}$. That is, instead of a simple linear relation between DGR and $Z$, the authors suggest a broken power-law. In \citet{ROMAN-DUVAL17}, the authors showed that the DGR changes by factors of 3 to 7 in the Magellanic Clouds, where metallicity is considered to be constant. This result also indicates a variable DTM. In \citet{GIANNETTI17}, the authors found a ${\rm DGR}(Z) \propto Z^{1.4}$ in a sample set composed by 23 massive and dense star-forming regions in the far outer MW.

% Why studying a single galaxy is meaningful
In this work, we revisit the possible variation of DTM in a single galaxy, M101. There are several benefits to studying DTM within a single galaxy. First, metallicity measurements are calibrated more uniformly within one galaxy than across galaxies, which is crucial for studying DTM variation \citep{REMY-RUYER14, BERG15, CROXALL16}. Moreover, focusing on one galaxy can avoid the problem in galaxy-integrated results that DTM can be underestimated by integrating over dust-poor HI in outer disks \citep{DRAINE07}. By comparing the DTM within one galaxy and across galaxies, we will also be able to determine whether the possible variation in DTM depends more on local physical properties or galactic properties. Lastly, observations within one galaxy would have the minimum differences of MW foreground, calibration, and background level estimation, which means the data are more uniform.

% Why M101 is a perfect target for this study
M101 is an ideal target for this study for four reasons: 1) M101 has one of the most detailed studies of its metallicity from the Chemical Abundances Of Spirals survey \citep[CHAOS,][]{BERG15, CROXALL16}, based on electron temperature ($T_e$) derived from auroral line measurements. 2) M101 has the largest metallicity gradient among those galaxies where direct $T_e$-based metallicity measurements are available, ranging $7.5 \la \metal \la 8.8$ \citep{CROXALL16}. This range covers both as high as the solar neighborhood and as low as the turning point in \citet{REMY-RUYER14} broken-power law. 3) M101 has a good radial resolution even at far-infrared (FIR) observations because it is nearby (distance $\sim \rm 6.7~Mpc$), physically large (the 25th magnitude isophote in B band, or \rtf, is $0.2^\circ=23.4~{\rm kpc}$ at distance $\rm 6.7~Mpc$), and relatively face on \citep[inclination $\approx 16^\circ$,][]{FREEMAN01, MAKAROV14}. 4) M101 also has high sensitivity \textsc{Hi} and CO maps \citep{WALTER08, LEROY09}, which let us map the total gas distribution.

% Paper layout
This paper is presented as follows. \secref{sec: observations} presents FIR, \textsc{Hi}, CO, and other supporting data used in this study, with our data processing procedures. The five modified blackbody (MBB) model variants and the fitting methodologies are described in \secref{Sec: methods}. We present our fitting results in \secref{sec: results}, and compare them with known physical limitations and statistical properties. In \secref{sec: discussions}, we discuss the implication of our results, and the relation between our DTM and previous findings. Finally, we give our conclusions in \secref{sec: conclusions}.

\section{Observations} \label{sec: observations}
\subsection{Data}\label{sec: data}
In this section, we introduce the multi-wavelength measurements of M101 from several surveys and their uncertainties, which we adopted for this study. The physical properties (position, distance and orientation) of M101 adopted for this study are listed in Table \ref{tab: samples}.
%%
% Tabulate M101 properties
\begin{deluxetable}{lll}
\tablecaption{Properties of M101. \label{tab: samples}}
\tablehead{\colhead{Property} & \colhead{Value} & \colhead{Reference}}
\startdata
R.A. (2000.0)	& 14h~03m~12.6s & (1) \\
Dec (2000.0)	& +54d~20m~57s & (1) \\
Distance		& 6.7~Mpc\tablenotemark{$\dagger$} &(2) \\
\rtf			& $0^\circ.19990$ & (1) \\
Inclination		& $16^\circ$ & (1) \\
P.A.			& $38^\circ$ &(3) \\
$\alpha_{{\rm CO}~J=(2-1)}$\tablenotemark{*}	& $\rm (2.9/R_{21})~M_\sun~pc^{-2} (K~km~s^{-1})^{-1}$ & (4) \\
$R_{21}$ & 0.7 & (4) \\
\enddata
\tablenotetext{\dagger}{Consistent with the value in \citet{SHAPPEE11}.}
\tablenotetext{*}{See \secref{subsec: CO} for discussion of the $\alpha_{\rm CO}$ factor we use.}
\tablerefs{(1) HyperLeda database (\url{http://leda.univ-lyon1.fr/}), \citet{MAKAROV14}; (2) \citet{FREEMAN01}; (3) \citet{SOFUE99}; (4) \citet{SANDSTROM13}.}
\end{deluxetable}

\subsubsection{Infrared Imaging} \label{subsec: IR}
We use FIR images from the ``Key Insights on Nearby Galaxies: A Far-Infrared Survey with \textit{Herschel}'' survey \citep[KINGFISH,][]{KENNICUTT11} to fit dust surface densities in M101. KINGFISH imaged 61 nearby galaxies in the FIR with the \textit{Herschel Space Observatory} \citep{PILBRATT10}, covering $70~\micron$, $100~\micron$, and $160~\micron$ from Photoconductor Array Camera and Spectrometer \citep[PACS,][]{POGLITSCH10}, and $250~\micron$, $350~\micron$, and $500~\micron$ from Spectral and Photometric Imaging Receiver \citep[SPIRE,][]{GRIFFIN10}. We do not include the $70~\micron$ flux in our SED modeling because stochastic heating from small dust grains makes non-negligible contribution in that spectral range \citep{DRAINE07}, which is not accounted for by the simple SED models we employ in this study. The PACS images were processed from level 1 with \texttt{Scanamorphos v16.9} \citep{ROUSSEL13} by the KINGFISH team. The SPIRE images were processed with \texttt{HIPE} \citep{OTT10} version \texttt{spire-8.0.3287} and from level 1 to final maps with \texttt{Scanamorphos v17.0} \citep{ROUSSEL13} by the KINGFISH team. According to the KINGFISH DR3 user guide \citep{KINGFISH13}, the SPIRE images have been multiplied by correction factors of 0.9282, 0.9351, and 0.9195 for SPIRE250, SPIRE350, and SPIRE500, respectively, due to improved effective beam size estimation. The FWHMs are approximately $7\arcsec.0 = 0.23~\rm kpc$, $11\arcsec.2 = 0.36~\rm kpc$, $18\arcsec.2 = 0.59~\rm kpc$, $24\arcsec.9 = 0.81~\rm kpc$, and $36\arcsec.1 = 1.17~\rm kpc$ for the 100\micron, 160\micron, 250\micron, 350\micron, and 500\micron\ band images, respectively.

\subsubsection{\textsc{Hi}} \label{subsec: HI} % THINGS & surface density
We obtain \textsc{Hi} 21 cm line data from ``The \textsc{Hi} Nearby Galaxy Survey'' \citep[THINGS,][]{WALTER08}. The images were obtained at the Very Large Array (VLA)\footnote{The VLA is operated by the National Radio Astronomy Observatory (NRAO), which is a facility of the National Science Foundation operated under cooperative agreement by Associated Universities, Inc.}. The M101 dataset in this survey has (10\arcsec.8, 10\arcsec.2)$\sim$(0.35~kpc, 0.33~kpc) angular resolution and $\rm 5.2~km~s^{-1}$ velocity resolution with natural weighting. The observed 21 cm emission can be converted to \textsc{Hi} column density ($N_{\rm HI}$) via Eq. (1) and Eq. (5) in \citet{WALTER08} assuming it is optically thin, and then further converted to surface density $\Sigma_{\rm HI}$ by multiplying by the atomic weight of hydrogen. The uncertainty in the THINGS survey is dominated by the estimated zero-point uncertainty in \textsc{Hi}, which is around $1~\rm M_\sun/pc^2$, corresponding to 0.04 to 0.17 dex in the center of M101 (molecular gas dominated region), 0.03 to 0.04 dex for most atomic gas dominated region, and goes above 0.08 dex for the outer most pixels.

\subsubsection{CO and Total Gas} \label{subsec: CO}
We obtain CO emission line measurements from the ``HERA CO Line Extragalactic Survey'' \citep[HERACLES,][]{LEROY09, SCHRUBA11, SCHRUBA12, LEROY13}, a survey mapping the ${\rm ^{12}CO}~J=(2-1)$ rotational line at $230.538~\rm GHz$ of 48 nearby galaxies, including M101. The observation was carried out with Heterodyne Receiver Array \citep[HERA,][]{SCHUSTER04} on the IRAM 30-m telescope\footnote{IRAM is supported by CNRS/INSU (France), the MPG (Germany) and the IGN (Spain).}. The survey has 13\arcsec\ angular resolution and $\rm 2.6~km~s^{-1}$ velocity resolution. The CO line integrated intensity can be converted to surface density of $\rm H_2$ plus He ($\Sigma_{\rm mol}$) by:
\begin{equation}
    \Sigma_{\rm mol} = \alpha_{\rm CO}\frac{I_{{\rm CO}~J=(2-1)}}{R_{21}},
\end{equation}
where $\alpha_{\rm CO}$ is the CO-to-$\rm H_2$ conversion factor, see Table \ref{tab: samples}. The standard $\alpha_{\rm CO}$ is quoted for $I_{{\rm CO}~J=(1-0)}$, thus, we convert the $I_{{\rm CO}~J=(2-1)}$ with a fixed line ratio\footnote{We adopt the $\alpha_{\rm CO}$ value from \citet{SANDSTROM13}, which the authors originally derived with $I_{{\rm CO}~J=(2-1)}$ data and convert with $R_{21}=0.7$. Thus we need to use the same $R_{21}$ for consistency.} $R_{21}=(2-1)/(1-0)=0.7$ \citep{SANDSTROM13}.

% Total gas
With $\Sigma_{\rm HI}$ and $\Sigma_{\rm mol}$, we calculate the total gas mass surface density ($\Sigma_{\rm gas}$) with Eq. \ref{eq: total gas}. A multiplier of value 1.36 is included in $\Sigma_{\rm mol}$ for helium mass \citep{SANDSTROM13}. We multiply the $\Sigma_{\rm HI}$ by this factor to calculate the total gas surface density correctly.:
\begin{equation} \label{eq: total gas}
\Sigma_{\rm gas} = 1.36~\Sigma_{\rm HI} + \alpha_{\rm CO}\frac{I_{{\rm CO}~J=(2-1)}}{R_{21}}
\end{equation}

We have checked that a metallicity dependent $\alpha_{\rm CO}$ \citep{WOLFIRE10, BOLATTO13} would make no significant difference in $\Sigma_{\rm gas}$ because in the region where H$_2$ is important in M101, the metallicity is still relatively high. See more discussion in \secref{sec: alpha_CO discussion}.

\subsubsection{Metallicity\label{subsec: metal_data}}
We obtained metallicity measurements from CHAOS survey \citep{CROXALL16}. Measurements were taken in 109 \textsc{Hii} regions by the Multi-Object Double Spectrographs (MODS) on the Large Binocular Telescope \citep[LBT,][]{POGGE10}. They derived $T_e$ from a three-zone model with \textsc{[Oiii]}, \textsc{[Siii]}, and \textsc{[Nii]} line ratios. The electron densities are derived from \textsc{[Sii]} line ratios. This gives us gas phase oxygen abundances in 74 \textsc{Hii} regions inside M101, and also an average metallicity gradient spread over the galactocentric radius considered in this study. We will compare our derived DGR with their derived metallicity gradient \citep[Eq. 10 in][second line\footnote{Instead of the 7.4 Mpc distance quoted in \citet{CROXALL16}, we used a galaxy distance of 6.7 Mpc, thus we multiplied the slope in their Eq. 10 by $\frac{7.4}{6.7}$ to account for the difference.}]{CROXALL16}.  The uncertainty in $\metal$ from the average metallicity gradient is $\sim0.02$ dex in the center and $\sim0.07$ dex in the outer most part.
% 2018/02/14 checked: both our D and R25 are different from Croxall+2016, thus the conversion is necessary.
% 2018/07/25 there are two formulae in Croxall+2016 Eq. 10, one in kpc and the other in r25. We used the one in kpc.
% In Croxall+2016, there is no correction factor for O/H. OI and OIV are neglected.

\subsubsection{Star formation rate and stellar mass\label{subsec: other}}
We calculate star formation rate surface density ($\Sigma_{\rm SFR}$) from the Galaxy Evolution Explorer (GALEX) FUV \citep{MARTIN05} and \textit{Spitzer} Multiband Imaging Photometer (MIPS) 24 \micron\ data \citep{WERNER04, RIEKE04}, and stellar mass surface density ($\Sigma_\star$) from \textit{Spitzer} Infrared Array Camera (IRAC) 3.6 \micron. These data are from the Local Volume Legacy survey \citep[LVL,][]{DALE09}.

We use the following equation to convert observed FUV and IR emission to $\Sigma_{\rm SFR}$:
\begin{equation} \label{eq: SFR}
\Sigma_{\rm SFR}=(8.1\times 10^{-2} I_{\rm FUV} + 3.2 \times 10^{-3} I_{24})\cos i,
\end{equation}
where $i$ is the inclination of M101. $\Sigma_{\rm SFR}$ is in $M_{\sun}~\rm kpc^{-2}~yr^{-1}$, and both $I_{\rm FUV}$ and $I_{24}$ are in $\rm MJy~sr^{-1}$. Eq. \ref{eq: SFR} is adopted from \cite{LEROY08}, and it is functionally similar to the prescription in \citet{KENNICUTT12}.

For converting 3.6 \micron\ SED to $\Sigma_\star$, we use the relation:
\begin{equation}
    \Sigma_\star = 350I_{3.6}\cos i,
\end{equation}
where $\Sigma_\star$ is in $M_{\sun}~\rm pc^{-2}$, and $I_{3.6}$ is in $\rm MJy~sr^{-1}$. Note that the appropriate mass to light ratio ($\Upsilon_\star^{3.6}$) remains a topic of research \citep{MCGAUGH14, MEIDT14}. Here, we assume the $\Upsilon_\star^{3.6}=0.5$ \citep{MCGAUGH14}, see discussions in \citet{LEROY08} and A. K. Leroy et al. (2018, in preparation).

\subsection{Data processing\label{sec: data proc}}
\subsubsection{Background subtraction} \label{subsec: background error}
% Background subtraction
The IR and GALEX images that we use include contributions from various backgrounds and foregrounds. Throughout this study, we will neglect the structure in MW foreground over the relatively small angular ($r_{25}=0^\circ.2$) extent of M101. To estimate the foreground/background (hereafter referred to as background) level for each image, we need a uniform definition of background region. We define our background region as where $N_{\rm HI} < 1.0\times \rm 10^{18}~cm^{-2}$. For the GALEX map, we take the mean value in the background region as recommended due to the Poisson statistics of the GALEX counts. For the IR images, we fit a tilted plane and iteratively reject outliers. This includes several steps: we fit a tilted plane to all the background pixels. We then subtract the tilted plane from the data and calculate the absolute deviation (AD) from the median for all pixels and derive median absolute deviation (MAD). Finally, we use only the pixels with AD smaller than three times MAD to fit a tilted plane, and iterate over step two and three for five times, keeping the last fitted tilted plane as the background to be removed.

After background subtraction and convolution (\secref{subsec: convolution}), we calculate the covariance matrix\footnote{A matrix with its i-j element as the i-band to j-band covariance. Our covariance matrix has a dimension of 5x5, corresponding to the 100$-$500 \micron\ bands in \textit{Herschel}.} in the background region of the five \textit{Herschel} bands. This covariance matrix ($\mathcal{C}_{\rm bkg}$) will play an important role in calculation of likelihood in our fitting procedure because it incorporates the observed band-to-band correlation in the noise due to confusion and other astronomical sources into our fitting (\secref{subsec: fitting}). 

\subsubsection{Convolution}\label{subsec: convolution}
Maps obtained from different surveys do not have the same pixel scale and point spread function (PSF). In order to compare them pixel-by-pixel, we first convolve all the maps to match the PSF of SPIRE500 using the \texttt{convolve\_fft} function in \texttt{astropy.convolution} \citep{ASTROPY13}. Most kernels in this study were adapted from \citet{ANIANO11}, except the Gaussian kernels for THINGS and HERACLES surveys. For these two surveys, we built elliptical or circular Gaussian kernels according to their beam sizes \citep{WALTER08, LEROY09} to convolve them to match a Gaussian PSF with 25\arcsec\ FWHM. Then, we convolve the images with a second kernel from \citet{ANIANO11}, which convolves Gaussian PSF with 25\arcsec\ FWHM to SPIRE500 PSF.

\subsubsection{Alignment\label{subsec: alignment}}
% Alignment
After convolution, we align the coordinates of all the images with the SPIRE500 image and its pixel scale using the function \texttt{reproject\_exact} in \texttt{reproject}, an \texttt{astropy} affiliated package. The final pixel scale is 14.0\arcsec, or $\rm \sim 0.45~kpc$, which is smaller than half of SPIRE500 PSF FWHM, 36\arcsec, thus enough for properly sampling the PSF. In the final images, one resolution element contains $\sim 5.2$ pixels, therefore, neighboring pixels are not independent.

% Voronoi binning
\subsubsection{Binning\label{subsec: voronoi}}
One of our main interests is to analyze DTM in regions with $\metal \la 8.0$, where the relation of DTM with metallicity is expected to change \citep{HIRASHITA99, HIRASHITA11, REMY-RUYER14}. However, individual pixels in the low metallicity region, or outer disk, tend to have insufficient signal-to-noise ratio (SNR) for analysis. One way we can solve this problem is to bin neighboring pixels together and average the measured quantities in those pixels to increase SNR according to:

\begin{equation}
    {\rm SNR_{avg}} = \frac{(\sum_i {\rm Signal}_i)/n}{\sqrt{(\sum_i {\rm Noise}_i^2)/n^2}},
\end{equation}

where the summation is over resolution elements inside the binned and $n$ is the number of resolution elements. As a consequence, uniform binning requires all regions on the map to sacrifice their spatial resolution in order to recover the regions with lower SNR, which means some structures that could have been resolved would be smoothed out in the binning process. To optimize the resolution and extend to the outer disk simultaneously, we choose to use adaptive binning: binning more pixels together in the low SNR region, while binning fewer pixels together or leaving pixels as individuals in the high SNR region.

\begin{figure}
\centering
\includegraphics[width=\columnwidth]{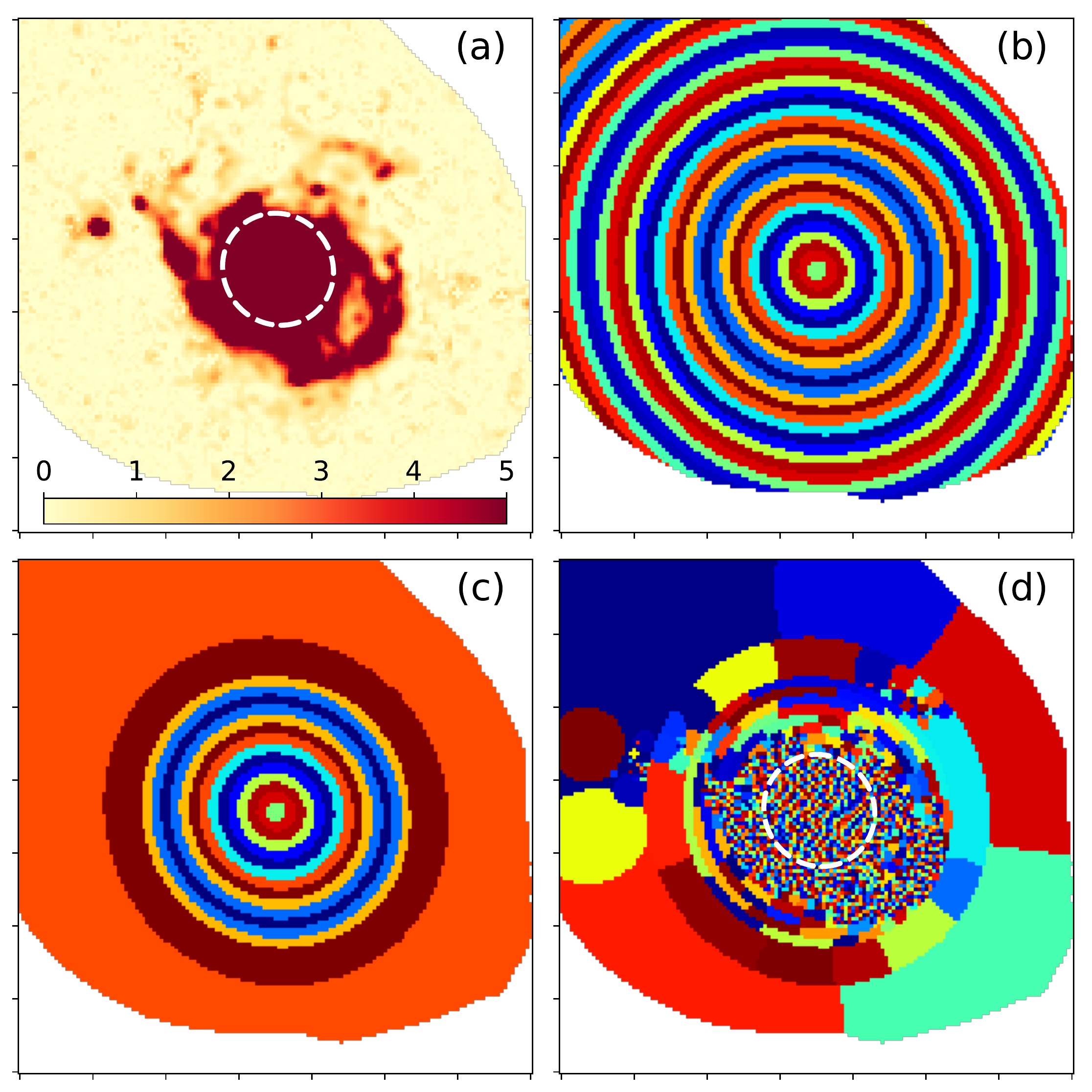}
\caption{Voronoi binning process in this study. (a) The worst-SNR map. Among the 16,403 points, 58\% have their worst SNR in PACS100. Both PACS160 and SPIRE500 take around 18\%. (b) The initial radial cut. (c) The final radial cut after grouping rings according to target SNR. (d) The final binned regions. The white circles in panel (a) and (d) show the radius 7.4 kpc. All pixels within 7.4 kpc remain unbinned.\label{fig: voronoi}}
\end{figure}

The adaptive binning method we choose is the \texttt{voronoi\_2d\_binning} function \citep{CAPPELLARI03}. Instead of directly apply the algorithm to the entire SED, we execute some extra procedures listed below in order to preserve radial information:
\begin{enumerate}
    \item We calculate SNR map for all five \textit{Herschel} bands using the square root of diagonal terms in the covariance matrix ($\mathcal{C}_{\rm bkg}$), which is the variance of each band, as the noise of each band.
    \item For each pixel, we select the lowest SNR among five bands at that pixel to build the worst SNR map, which is plotted in Figure \ref{fig: voronoi} (a). This worst SNR map is used for the subsequent binning process in order to make sure all five bands will reach the target SNR with the same binned regions. 58\% of pixels have their worst SNR from PACS100.
    \item We cut the target galaxy into concentric rings with the same radial spacing, which is set to be the same as the FWHM of the SPIRE500 PSF. This initial radial cut is shown in Figure \ref{fig: voronoi} (b).
    \item Starting from the outermost ring, if the average SNR of all pixels within a ring is lower than target SNR, we combine it with one ring inside until target SNR is achieved. This final radial cut is shown in Figure \ref{fig: voronoi} (c). The target SNR is set to be 5. However, since the pixels are oversampled with the SPIRE500 PSF (see \secref{subsec: alignment}), the effective target SNR is $ 5 / \sqrt{5.2} \sim 2.2 $.
    \item We apply \texttt{voronoi\_2d\_binning} with \texttt{targetSN} set to 5, to each ring from Step 4 and worst SNR map from Step 2 to generate the final binned regions, as shown in Figure \ref{fig: voronoi} (d).
\end{enumerate}
Note that we discard the \texttt{roundness} threshold in the original function \citep{CAPPELLARI03}. This \texttt{roundness} threshold makes sure all binned region are nearly circular, which will result in malfunctions when we cut the image into concentric circles at the beginning. All pixels within radius 7.4 kpc ($0.3~\rm r_{25}$) have high enough SNR thus remain unbinned.

\section{Methods\label{Sec: methods}}
\subsection{Models\label{Sec: Model}}
In this work, we focus on the FIR part of the dust emission SED. It is reasonable to assume that emission from dust grains in thermal equilibrium dominates the FIR range \citep{LI01, BLAIN02, GORDON14}, therefore we start with fitting the FIR emission with a modified blackbody (MBB) model:
\begin{equation}
I_\nu = \kappa_{\nu}\Sigma_d B_\nu(T_d),
\end{equation}
where $I_\nu$ is the specific intensity, $\kappa_{\nu}$ is the wavelength-dependent emissivity, $\Sigma_d$ is the dust surface density, and $B_\nu(T_d)$ is the blackbody spectral radiance at dust temperature $T_d$. An empirical power law emissivity is often assumed, that is, $\kappa_\nu = \kappa_{\nu_0}(\nu/\nu_0)^\beta$, where the emissivity index $\beta$ is a constant and $\nu_0=c/\lambda_0$. Throughout this study, $\lambda_0=160~\micron$ is used.

There are a few possible drawbacks to this simple model, some of them are physical, and the others are inherent to the process of fitting the model. The physical drawbacks include: 1) The simple model above does not allow for wavelength or environmental dependence of $\beta$, which might exist \citep{REACH95, FINKBEINER99, LI01, GORDON14}. 2) The model does not include stochastic heating \citep{DRAINE07}, which might contribute to our shortest wavelength observation due to the width of the response functions of the PACS instruments. 3) The model does not include the broadening in the SED due to multiple heating conditions involved in one resolution element \citep{DALE01}. The fitting process drawbacks include: 1) $\kappa_{\nu_0}$ and $\Sigma_d$ are completely degenerate, thus there will be an inherent uncertainty in $\Sigma_d$ from how we determine the $\kappa_{\nu_0}$ value. 2) Due to the nature of this model, $\beta$ and $T_d$ are covariant, since they both shift the peak wavelength of the SED. Thus, there might be artificial correlation between them. \citet{KELLY12} demonstrated this artificial correlation with traditional $\chi^2$-minimization fitting.

We calibrate $\kappa_{\nu_0}$ with high-latitude MW diffuse ISM following the approach in \citet{GORDON14} (see \secref{Sec: calibration}). It is possible that this calibration is not appropriate at all local environmental conditions and it would result in a systematic uncertainty in our results (see \secref{subsec: discuss_emissivity} for further discussion). We also use a probabilistic fitting procedure following \citet{GORDON14} that lets us assess the correlations between fit parameters and properly marginalize over the degeneracy between $\beta$ and $T_d$. Still, there is no simple way to solve all the physical drawbacks of the MBB model. In order to address the physical shortcomings of the MBB model, we construct five variant models. These each address a shortcoming of the MBB. They are not all mutually exclusive, and a full model \citep[e.g.,][]{DRAINE07} might incorporate several of these. Our goal here is to identify the simplest possible modifications that yield a good fit to the IR SED. These variants are listed below:

\subsubsection{Simple emissivity (SE)\label{subsubsec: SE}}
Here, we assume a simple power-law emissivity, which gives a dust emission SED described by the following equation:
\begin{equation} \label{eq: SE}
I_\nu = \kappa_{\nu_0}(\frac{\nu}{\nu_0})^\beta \Sigma_d B_\nu(T_d).
\end{equation}
The free parameters in this model are $\Sigma_d$, $T_d$ and $\beta$. This method allows $\beta$ to vary spatially, thus could partially avoid the environmental-dependent $\beta$ drawback. However, it is also heavily affected by the possible artificial correlation between $\beta$ and $T_d$.

\subsubsection{Fixing $\beta$ (FB)\label{subsubsec: FB}}
Using the same functional form as Eq. \ref{eq: SE}, we can also fix the $\beta$ value. This is one way to remove the inherent covariance between $T_d$ and $\beta$ based on what is expected for the optical properties of ISM dust grain materials. In some previous studies \citep{MENNELLA98, BOUDET05, GALLIANO17} and our preliminary test of SE method, there are fitting results with anti-correlated $T_d$ and $\beta$. This could mean that $\beta$ is a function of $T_d$, however, due to the degeneracy of $T_d$ and $\beta$ in the model, it is also possible that this anti-correlation is all, or partially, artificial \citep{SHETTY09, SHETTY09B, KELLY12}. In the latter case, fixing $\beta$ can improve the accuracy of fitted $T_d$ \citep{SHETTY09B}. Thus, we adapted $\beta=2$ from previous studies \citep{REACH95, DUNNE01, DRAINE07} as a variation of MBB spectrum. We also tested $\beta$ values of 1.6, 1.8, and 2.2 and the difference in $\Sigma_d$ and chi-square values between them and $\beta=2$ results are insignificant. The insensitivity of the resulting $\Sigma_d$ to our choice of $\beta$ results from the fact that we calibrate the emissivity for each $\beta$ value accordingly. The process of emissivity calibration is described in \secref{Sec: calibration}. It is also true for the other methods where we also have $\beta$ fixed at 2 at short wavelength or the whole spectral range.

\subsubsection{Broken Emissivity (BE)}
It is possible that the dust emissivity is not a simple power law, but varies with wavelength. Previous studies have shown that the emissivity in the long wavelength end tends to be flatter than the short wavelength end. Thus, many authors including \citet{REACH95} and \citet{GORDON14} have tried to build more complicated forms of emissivity as a function of wavelength. Here, we adapted the BEMBB model in \citet{GORDON14}: assuming $\beta$ as a step function in wavelength, which makes the emissivity a broken-power law (Eq. \ref{eq: BE}).
\begin{equation}\label{eq: BE}
\kappa_\nu=\left\{\begin{array}{ll}
    \kappa_{\nu_0}(\frac{\nu}{\nu_0})^{\beta} & \lambda < \lambda_c \\
    \kappa_{\nu_0}(\frac{\nu_c}{\nu_0})^{\beta}(\frac{\nu}{\nu_c})^{\beta_2} & \lambda \geq \lambda_c
    \end{array}\right.
\end{equation}
$\lambda_c$ is the critical wavelength corresponding to the break, and $\nu_c$ is the frequency corresponding to $\lambda_c$. $\lambda_c$ is fixed at 300~\micron\ in this study. We explored varying the break wavelength with the spectral range of 50 to 600 \micron and found it had no major impact on the results. $\beta_2$ is the dust emissivity index at long wavelength. The short wavelength dust emissivity index $\beta$ is fixed at 2 in this study.

\subsubsection{Warm dust component (WD)}
In the spectral region below $100~\micron$, it is possible that the SED is affected by stochastic emission from small grains \citep{DRAINE07}, which is within the effective bandpass of the PACS100 response function (around $80$ to $120~\micron$). In this model, we add a second MBB component with $T_d=\rm 40~K$ to our SED, called ``warm dust'', to simulate the contribution from stochastically heated dust. We made this choice of $T_d$ to have the peak of warm dust SED at the boundary of PACS100 response function. The fraction of warm dust relative to total dust is symbolized as $f_W$. The fitting model in this method becomes (Note that both components have power-law emissivity with $\beta=2$):
\begin{equation}
I_\nu = \kappa_{\nu_0}(\frac{\nu}{\nu_0})^\beta\Sigma_d \Big((1-f_W) B_\nu(T_d) + f_W B_\nu(40K) \Big).
\end{equation}

To properly take this effect into account, one would need to adopt a complete physical dust model. However, among the dust properties, we are mainly interested in $\Sigma_d$, which is necessary for calculating DGR and DTM, and which does not require adopting a full dust model. This is because within our current understanding of dust heating and the dust grain size distribution, only a small fraction of the dust mass is stochastically heated \citep{DRAINE07}. Our preliminary test confirms this: the mass fraction of stochastically heated dust in the WD modeling is usually under 1\%. This means that we can still acquire reasonable accuracy in $\Sigma_d$ even when the SED of stochastically heated dust is not modeled with high accuracy. 

\subsubsection{Power Law distribution (PL)}\label{sec: PL}
At the SPIRE500 resolution, the FWHM of PSF would have a large physical size ($\sim 1.22~{\rm kpc}$). Thus, it is likely that there are various dust heating conditions within one resolution element. To attempt to model such a distribution of heating conditions, we adopt a model wherein a fraction ($1-\gamma$) of the dust mass is heated by a single value ISRF $U_{\rm min}$, while the other $\gamma$ fraction is heated by a distribution of ISRF between $U_{\rm min}$ and $U_{\rm max}$ with $\frac{d\Sigma_d}{dU}\propto U^{-\alpha}$ \citep{DALE01, DRAINE07}. Each mass fraction emits a FB MBB spectrum, which makes the total emission\footnote{The normalization factor $\frac{1-\alpha}{U_{\rm max}^{1-\alpha} - U_{\rm min}^{1-\alpha}}$ in Eq. \ref{eq: PL} only works when $\alpha \neq 1$. For $\alpha = 1$ (which is excluded in this study), one should use $\frac{1}{\ln(U_{\rm max}/U_{\rm min})}$ instead.}:
\begin{equation} \label{eq: PL}
\begin{array}{ll}
I_\nu = \kappa_{\nu_0}(\frac{\nu}{\nu_0})^\beta\Sigma_d & \Big((1-\gamma) B_\nu(U_{\rm min}) + \\
& \gamma \frac{1-\alpha}{U_{\rm max}^{1-\alpha} - U_{\rm min}^{1-\alpha}}\int^{U_{\rm max}}_{U_{\rm min}}U^{-\alpha}B_\nu(U)dU \Big).
\end{array}
\end{equation}
To calculate the equivalent MBB temperature, we convert $U$ to $T_d$ as $U \propto T_d^{\beta + 4}$, with a normalization of $U=1$ corresponding to $T_d=\rm 18~K$ \citep{DRAINE14}. This approach adds several free parameters, however, since we do not have good constraints for all of them, we fix some parameters before fitting: $U_{\rm max}$ is fixed at $10^7$ \citep[following][]{ANIANO12}, and $\beta$ is fixed at 2. Thus, the number of free parameters is 4, which is not a major difference from the other models.

\subsection{Fitting techniques} \label{subsec: fitting}
% Tabulate grid space definition
\begin{deluxetable}{lllll}
\tablecaption{Grid parameters for fitting.\label{tab: grid space}}
\tablecolumns{4}
\tablewidth{0pt}
\tablehead{
\colhead{Parameter} &
\colhead{Range} &
\colhead{Spacing} &
\colhead{Range$_c$\tablenotemark{f}} &
\colhead{Spacing$_c$}}
\startdata
$\logt\Sigma_d$ & -4 to 1\tablenotemark{a} & 0.025 & $\pm0.2$ & 0.002\\
$T_d$ & 5 to 50\tablenotemark{b} & 0.5 & $\pm1.5$ & 0.1\\
$\beta$ & -1.0 to 4.0\tablenotemark{c} & 0.1 & $\pm0.3$ & 0.02\\
$\lambda_c$ & 300\tablenotemark{d} & N/A & 300 & N/A\\
$\beta_2$ & -1.0 to 4.0 & 0.25 & $\pm0.3$ & 0.02\\
$f_W$ & 0.0 to 0.05 & 0.002 & $\pm0.006\tablenotemark{g}$ & 0.0005\\
$\alpha$ & 1.1 to 3.0 & 0.1 & $\pm0.3$ & 0.01\\
$\logt\gamma$ & -4.0 to 0.0 & 0.2 & $\pm0.3$ & 0.1\\
$\logt U_{\rm min}$ & -2.0 to 1.5\tablenotemark{e} & 0.1 & $\pm0.1$ & 0.01\\
$\logt U_{\rm max}$ & 7 & N/A & 7 & N/A \\
\enddata
\tablecomments{(a) $\Sigma_d$ in $M_\sun ~{\rm pc}^{-2}$. (b) In K. (c) For SE only. All the others are fixed at $\beta=2$. (d) In \micron. (e) $9.3\leq T_d \leq 35.6~\rm K$ under our conversion. (f) Range for second iteration during calibration. (g) While none negative.}
\end{deluxetable}
% Parameter space & grid fitting method
We follow the fitting techniques in \citet{GORDON14}: we build model SEDs on discrete grids in parameter space, and then calculate the likelihood for all models given the SED in each binned region. The multi-dimensional (3 dimensional for SE, BE and WD methods, 2 for FB and 4 for PL) grids have axes defined in \secref{Sec: Model}, and grid spacing defined in Table \ref{tab: grid space}.

% Band integration
For each grid point, we can generate a model SED $M_{ij...d}(\nu)$, where the subscript represents a unique combination of parameters in the grid with $d$ dimensions. The calculated model is a continuous function of frequency $\nu$. To compare with the real observation, we integrated $M_{ij...d}(\nu)$ over the response function $R^n(\nu)$ of each band $n$ in PACS and SPIRE with the following integral:
\begin{equation}
\overline{M^n_{ij...d}} = \frac{\int^\infty_0 R^n(\nu)M_{ij...d}(\nu)d\nu}{\int^\infty_0 R^n(\nu)(\nu_n/\nu)d\nu}
\end{equation}
Note that the denominator is added to account for the fact that \textit{Herschel} intensities are quoted assuming a spectrum with $S(\nu) \propto \nu^{-1}$ within the response function. The $\nu_n$ values are the frequencies corresponding to the representative wavelength at each band, that is, 100, 160, 250, 350, and 500 \micron.

% Likelihood function
Next, in each binned region, we calculate the relative likelihood ($\mathcal{L}$) of the model SED ($\overline{M_{ij...d}}$) given the observed SED ($I_{\rm obs}$) assuming Gaussian errors\footnote{See \citet{GORDON14} for discussion about statistical advantages of this matrix form definition}, that is:
\begin{equation}\label{eq: likelihood}
\mathcal{L}(\overline{M_{ij...d}}|I_{\rm obs}) = \exp\big( -\frac{1}{2} \chi^2_{ij...d} \big),
\end{equation}
where
\begin{equation}\label{eq: chi2}
\chi_{ij...d}^2 \equiv  (\overline{M_{ij...d}}-I_{\rm obs})^T \mathcal{C}^{-1} (\overline{M_{ij...d}}-I_{\rm obs})
\end{equation}
and
\begin{equation}\label{eq: covariance_sum}
\mathcal{C} = \mathcal{C}_{\rm bkg} + \mathcal{C}_{\rm cal}.
\end{equation}
The $^T$ sign represents the transpose matrix, and $^{-1}$ sign represents the inverse matrix. $\mathcal{C}_{\rm bkg}$ is the background covariance matrix discussed in \secref{subsec: background error} with values:
\begin{eqnarray}
\mathcal{C}_{\rm bkg} = 
 \left[
\arraycolsep=\colspace
\begin{array}{ccccc}
    1.548 & 0.09 & 0.057 & 0.025 & 0.01 \\
    0.09 & 0.765 & 0.116 & 0.079 & 0.04 \\
    0.057 & 0.116 & 0.098 & 0.071 & 0.037 \\
    0.025 & 0.079 & 0.071 & 0.063 & 0.033 \\
    0.01 & 0.04 & 0.037 & 0.033 & 0.028 \\
\end{array}\right].
\end{eqnarray}
As described in \secref{subsec: voronoi}, $\mathcal{C}_{\rm bkg}$ will be lower for resolution elements binned together. For a binned region with a number of pixels greater than one resolution element (5.2 pixels, see \secref{subsec: alignment}), $\mathcal{C}_{\rm bkg}$ is divided by number of resolution elements in the region.

% Calibration errors
$\mathcal{C}_{\rm cal} = I^T \mathcal{M}_{\rm fit} I$ is the covariance matrix generated from calibration error, where $\mathcal{M}_{\rm fit}$ is the percentage calibration errors and $I$ is the observed SED at the binned region. There are two kinds of errors from calibration. The first one is absolute calibration uncertainty, estimated from the systematic uncertainty by comparing the calibrator to model \citep{BENDO17}. We assume this absolute calibration uncertainty will affect all the bands calibrated together at the same time, thus we will fill this uncertainty both in the diagonal terms and the band-to-band off diagonal terms in $\mathcal{M}_{\rm fit}$. The second one is the relative uncertainty, or random uncertainty, which is estimated from the ability of an instrument to reproduce the same measurement \citep{BENDO17}. We assume this noise is band-independent thus we only put it in diagonal terms in $\mathcal{M}_{\rm fit}$.

Among the \textit{Herschel} observations, the SPIRE instruments were calibrated with Neptune, and were estimated to have 4\% absolute calibration and 1.5\% relative calibration uncertainty. The PACS instruments were calibrated with 5 stars. and the result gave a 5\% absolute uncertainty and 2\% relative uncertainty \citep{PACS13, BALOG14}. In the diagonal terms in $\mathcal{M}_{\rm fit}$, where we need to consider both kinds of uncertainties, it is recommended that we should take the direct sum of the two errors instead of quadratic sum \citep{BALOG14, BENDO17}. Since our object is an extended source, we must also take the uncertainty in the beam shape into account when calculating calibration errors \citep{BENDO17}. It is recommended that we double the absolute uncertainties for this \citep{GORDON14}. The final $\mathcal{M}_{\rm fit}$ is:
\begin{eqnarray}\label{Eq: Mcal}
\mathcal{M}_{\rm fit} = 
 \left[
\arraycolsep=\colspace
\begin{array}{ccccc}
    \cpr & \cpa & 0 & 0  & 0 \\
    \cpa & \cpr & 0 & 0  & 0 \\
    0 & 0 & \csr & \csa & \csa \\
    0 & 0 & \csa & \csr & \csa \\
    0 & 0 & \csa & \csa & \csr \\
\end{array}\right].
\end{eqnarray}

\begin{figure*}
\centering
\includegraphics[width=\textwidth]{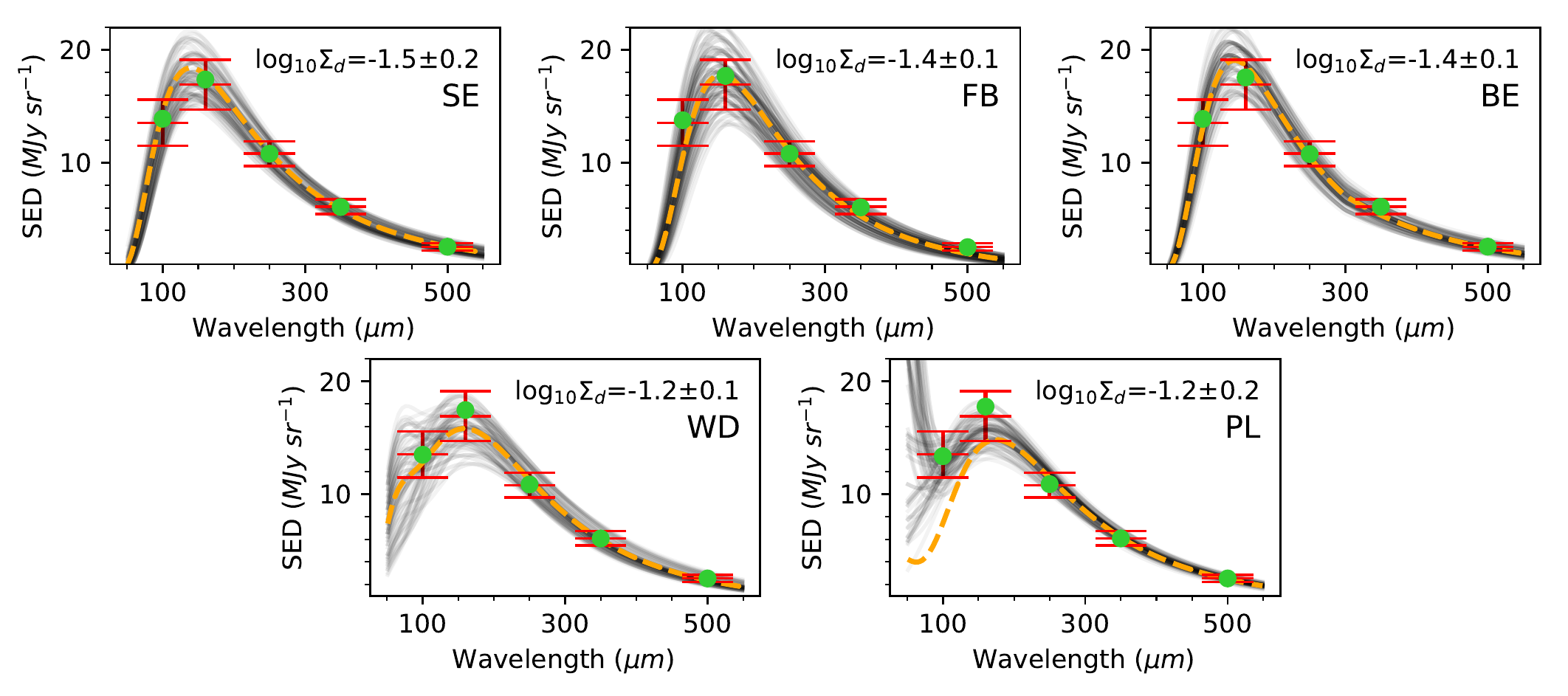}
\caption{An example of observed SED versus fitted SED from a single binned region. Red: The observed SED and error used in the fit. The error bars only include the square root of diagonal terms from the complete covariance matrix $\mathcal{C}$. Green dot: The SED convolved with response function. Orange dashed line: The model SED generated from expectation values in the fit. Gray lines: Some selected models with transparency proportional to $\mathcal{L}$. For each method, we randomly select 50 models from the subset $\mathcal{L}(M^n_{ij...d}|I^n) \geq max\big(\mathcal{L}(M^n_{ij...d}|I^n)\big)/1000$ for plotting. Note that both WD and PL methods allow FB components with peak wavelength below 100 \micron\, where we do not include observational constraint in this study. Therefore, the unusual shape in SED at short wavelength will not affect the fitting qualities of those models. However, we can still get similar expectation values in $\Sigma_d$ from these methods.\label{fig: example_model_merged}}
\end{figure*}
\begin{figure}
\centering
\includegraphics[width=\columnwidth]{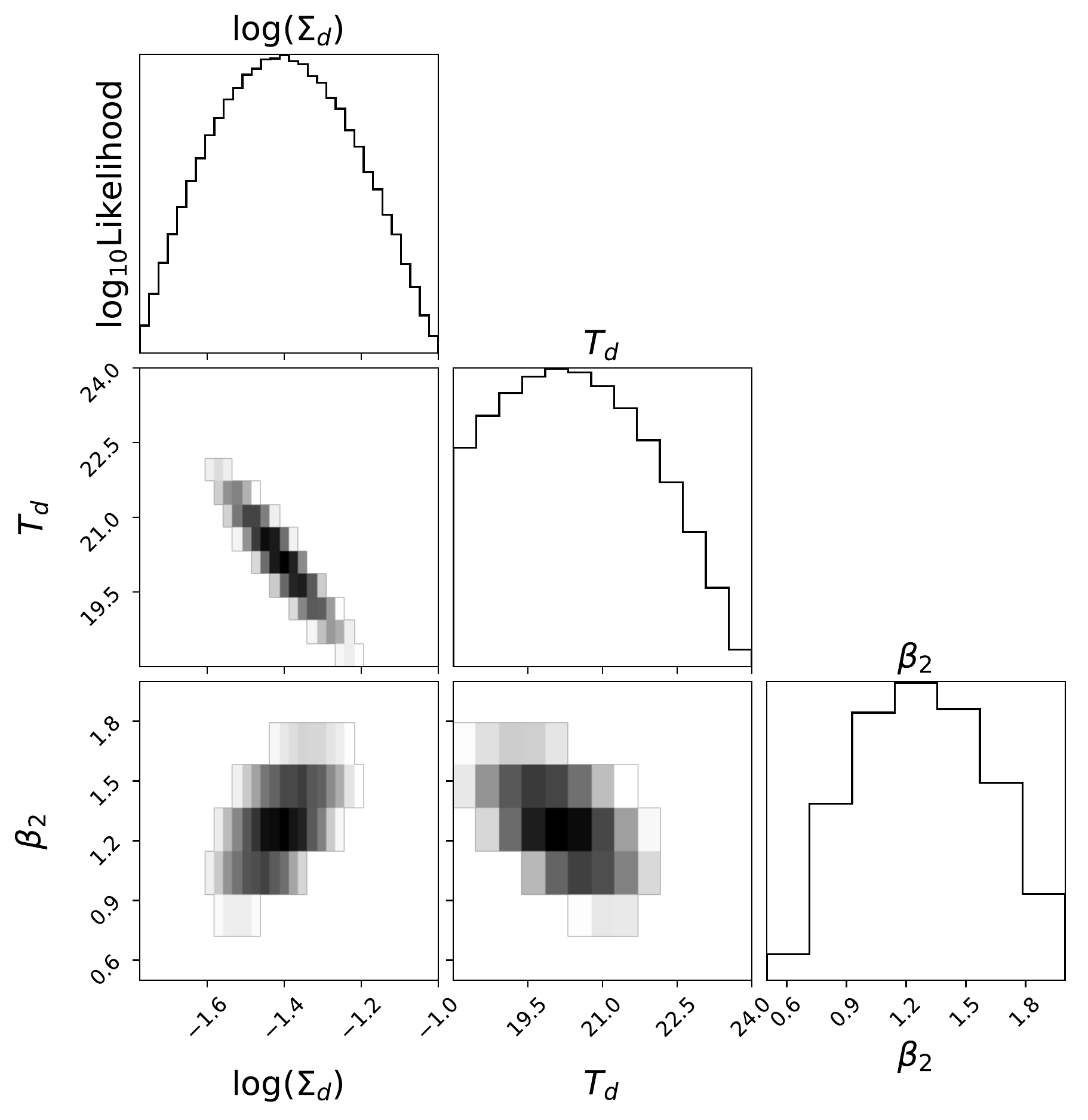}
\caption{Likelihood distribution in the parameter space from results of BE method at the same binned region in Figure \ref{fig: example_model_merged}. Both the histograms and 2-dimensional histograms are shown in log scale. The figure does not include the whole parameter space. It is magnified to emphasize the region with $\chi^2 \leq \big(min(\chi^2) + 6\big)$.\label{fig: example_model_corner}}
\end{figure}

With the relative likelihood $\mathcal{L}(\overline{M_{ij...d}}|I_{\rm obs})$ calculated, we can construct the full probability distribution function (PDF) for each parameter by summing over all other dimensions in parameter space. For example, if the index $i$ corresponds to $\Sigma_d$, then the PDF of $\Sigma_d$ with observed $I^n$ would be $P_{\Sigma_{d,i}} = \sum_{j...d} \mathcal{L}(\overline{M_{ij...d}}|I_{\rm obs})$. We can then calculate the expectation value\footnote{When calculating the expectation values, we use logarithmic scales for variables with logarithmic spacing in the grid.}, and the probability weighted 16\% and 84\% values, which represent the 1-$\sigma$ confidence interval and are sampled to represent the uncertainty of the fit. An example of observed SED versus fitted models with all methods is shown in Figure \ref{fig: example_model_merged}. An example of the log-scale likelihood distribution and correlation between fitting parameters is shown in Figure \ref{fig: example_model_corner}.

\subsubsection{Calibrating $\kappa_{160}$ \label{Sec: calibration}}
We use the procedure and integrated dust SED of the MW diffuse ISM from \citet{GORDON14} to calibrate $\kappa_{160}$ in our models. The SED was originally measured with Cosmic Background Explorer (COBE), where the $\lambda \geq 127~\micron$ measurements are from Far Infrared Absolute Spectrophotometer (FIRAS) and the 100 \micron\ measurement is from Diffuse Infrared Background Experiment (DIRBE). The resulting SED is 0.6887, 1.4841, 1.0476, 0.5432, and 0.2425 ${\rm MJy~sr^{-1}~(10^{20}~H~atom)^{-1}}$ for the 100, 160, 250, 350, and 500 \micron\ bands. These values differ from those given by \citet{GORDON14} because we include a factor of 0.97 for the molecular cloud correction \citep{COMPIEGNE11}. The ionized gas factor in \citet{COMPIEGNE11} is excluded because we do not include ionized gas through out this study, including the calculation of average DGR in the MW diffuse ISM \citep{JENKINS09, GORDON14}. The dust-to-Hydrogen mass ratio appropriate for this high-latitude diffuse region is calculated by averaging the depletion strength factor $\rm F_\star$ value over sightlines in \citet{JENKINS09} with similar hydrogen column densities as the observed region. The resulting $\rm F_\star$ is 0.36, and the dust-to-Hydrogen mass ratio is $1/150$, which corresponds to a dust surface density to H column density ratio of $5.30\times10^{-3}~{\rm M_\sun~ pc^{-2}}~(10^{20}~{\rm H~atom})^{-1}$.

During calibration, it is important to use the same models and fitting methods as the real fitting \citep{GORDON14}. We follow the same steps of our fitting techniques except four necessary differences: 1) We replace the original $\mathcal{M}_{\rm fit}$ with $\mathcal{M}_{\rm cali}$ (Eq. \ref{eq: M_cal cali}) for calibration since the calibration data came from COBE instead of \textit{Herschel}. Following \citet{FIXSEN97}, we assume 0.5\% relative uncertainty and 2\% absolute uncertainty for FIRAS (calibrating PACS160 and SPIRE bands), and 1\% relative uncertainty and 10\% absolute uncertainty for DIRBE (calibrating PACS100 \micron).
\begin{eqnarray}\label{eq: M_cal cali}
\mathcal{M}_{\rm cali} = 
 \left[
\arraycolsep=\colspace
\begin{array}{ccccc}
    \cdd & 0 & 0 & 0  & 0 \\
    0 & \cdf & \ccf & \ccf & \ccf \\
    0 & \ccf & \cdf & \ccf & \ccf \\
    0 & \ccf & \ccf & \cdf & \ccf \\
    0 & \ccf & \ccf & \ccf & \cdf \\
\end{array}\right]
\end{eqnarray}
2) No $\mathcal{C}_{\rm bkg}$ term is applied. $\mathcal{C}_{\rm cal}$ is the only variance term considered. 3) Due to the small uncertainty of COBE data, the normal parameter spacing is not finely-sampled enough to resolve the PDF for all the parameters. Thus, we use a two-step calibration: first, we fit with the normal parameter space; then reduce the parameter range to a smaller region near the peak with a finer spacing (see ``Range$_c$'' and ``Spacing$_c$'' columns in Table \ref{tab: grid space}); last, we fit with this new parameter spacing and report the results. 4) Our SED per hydrogen atom of the MW diffuse ISM is weaker than the one in \citet{GORDON14} by a factor of 0.97 due to the molecular cloud fraction.

% Tabulate calibration results
\begin{deluxetable*}{llcc}
\tablecaption{Results of calibrating emissivity to the MW high latitude SED.\label{tab: cali result}}
\tablecolumns{4}
\tablewidth{0pt}
\tablehead{
\colhead{Model} &
\colhead{$\kappa_{160}~({\rm cm^2~g^{-1}})$} &
\colhead{Other parameters} &
\colhead{Expectation values}}
\startdata
SE & $10.10\pm1.42$ & ($T_d$, $\beta$) & ($20.90\pm0.62$~K, $1.44\pm0.08$) \\
FB & $25.83\pm0.86$ & ($T_d$) & ($17.13\pm0.12$~K) \\
BE & $20.73\pm0.97$ & ($T_d$, $\beta_2$) & ($18.02\pm0.18$~K, $1.55\pm0.06$) \\
WD & $27.46\pm1.14$ & ($T_d$, $f_W$) & ($16.60\pm0.25$~K, $0.00343\pm0.00143$) \\
PL & $26.60\pm0.98$ & ($\alpha$, $\logt\gamma$, $\logt U_{\rm min}$) & ($1.69\pm0.19$, $-1.84\pm0.21$, $-0.16\pm0.03$)\\
\enddata
\end{deluxetable*}
The calibrated $\kappa_{160}$ values range from 10.48 to 21.16$~\rm cm^2~g^{-1}$, see complete results in Table \ref{tab: cali result}. This is a fairly large range, which indicates that the choice of model does affect the measurement of dust properties. Our results are comparable with calculated $\kappa_{160}$ values in literature, e.g., the widely used \citet{DRAINE07} model, with updates in \citet{DRAINE14}, gives $\kappa_{160}$ equal to 13.11$~\rm cm^2~g^{-1}$ for silicates and 10.69$~\rm cm^2~g^{-1}$ for carbonaceous grains, and 12.51$~\rm cm^2~g^{-1}$ in the combined model. The standard model in \citet{GALLIANO11} gives a value of 14$~\rm cm^2~g^{-1}$, and 16$~\rm cm^2~g^{-1}$ after replacing graphite with amorphous carbons. A recent calculation by \citet{RELANO18}, following the \citet{DESERT90} dust model, gives an equivalent $\kappa_{160}=22.97~\rm cm^2~g^{-1}$.

In the MBB model calibration process in \citet{GORDON14} and \citet{GORDON17}, the resulting $\kappa_{160}$ falls between 30.2 and 36.4$~\rm cm^2~g^{-1}$, depending on the model used. The common model between us is the SMBB in \citet{GORDON14}, where they have $\kappa_{160}=30.2~\rm cm^2~g^{-1}$, and our SE, where we have $\kappa_{160}=10.1~\rm cm^2~g^{-1}$. Our calibration method differs from \citet{GORDON14} in four ways: 1) With the values of COBE uncertainty we quote, we are allowed to have more deviation at 100 \micron\ than the other bands. On the other hand, \citet{GORDON14} have both correlated and uncorrelated uncertainty values uniform for all bands. 2) We use a $M_{\rm cali}$ which assumes 100 \micron\ calibration independent of the other bands since DIRBE and FIRAS were calibrated independently. \citet{GORDON14} assumed that all bands are correlated with the same absolute uncertainties. 3) We use a two-step fitting to increase the accuracy only for calibration, while \citet{GORDON14} used exactly the same methods for calibration and fitting. 4) Our SED per hydrogen atom of the MW diffuse ISM is weaker by a factor of 0.97 due to the molecular cloud fraction. In Section~\ref{sec:sensitivity} we discuss the sensitivity of the results to choices in the SED fitting and calibration in more detail.

\section{Results}\label{sec: results}
We fit the SEDs from all binned regions with all five MBB variants introduced in \secref{Sec: Model}. We calculate the DGR in each bin from the observed $\Sigma_{\rm gas}$ and the fitting results of $\Sigma_d$. Here, we look at the DGR and dust temperature radial gradients for each model, and at the residuals and reduced chi-square values about the best fit. Doing so, we will be interested in which models meet our physically motivated expectations and which models provide good fits to the SED. The complete fitting results are shown in Appendix \ref{app: fitting}, along with their correlations in Appendix \ref{app: corner}.

\subsection{DGR-metallicity relation\label{sec: max DGR}}
\begin{figure}
\centering
\includegraphics[width=\columnwidth]{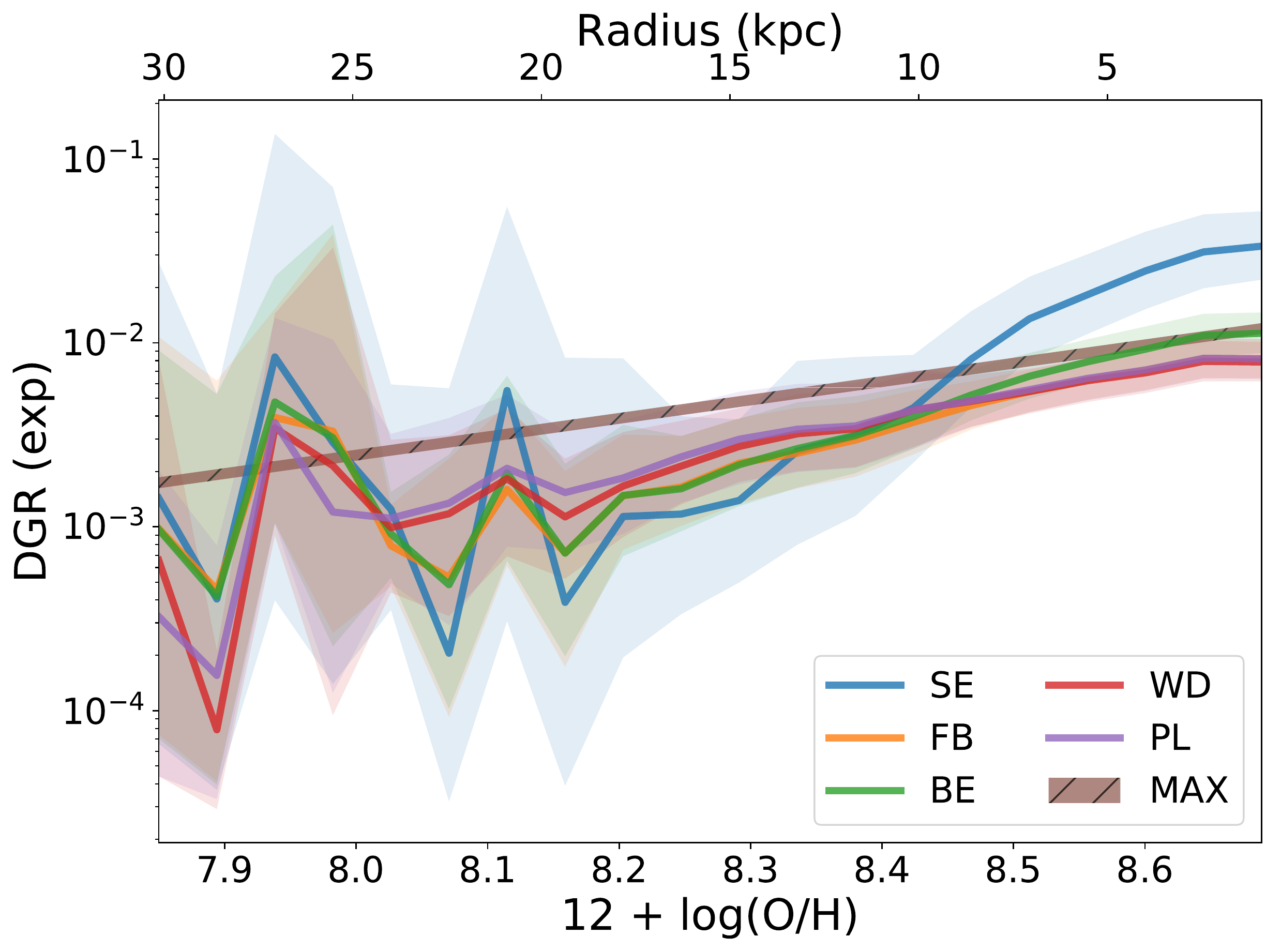}
\caption{DGR expectation values versus radius and metallicity. The shaded regions show the intrinsic scatter of DGR from $\Sigma_d$ fitting results and the zero-point fluctuation of $\Sigma_{\rm gas}$ (\secref{subsec: HI}). MAX is the maximum possible DGR calculated as a function of metallicity. The range is set by the difference between \citet{LODDERS03} and \citet{ASPLUND09} chemical composition, which is small at this plotting scale.\label{fig: DGR_grad_models_merged}}
\end{figure}
In Figure \ref{fig: DGR_grad_models_merged}, we plot the DGR-metallicity relation from all fitting methods. The metallicity-radius relation is calculated with Eq. 10 in \citet{CROXALL16}. We first separate M101 into 20 radial regions, and, at each region with $r_i\leq r < r_j$, we take the sum of the expectation value of dust mass divided by the total gas mass as the expectation value of DGR (${\rm < DGR >}$) in that region, that is:
\begin{equation}
{\rm <DGR>}_{ij} = \frac{\sum_{r_i\leq r_k < r_j}<\Sigma_d>_k A_k}{\sum_{r_i\leq r_k < r_j}M_{\rm gas,k}},
\end{equation}
where $<\Sigma_d>_k$ and $A_k$ are the expectation value of $\Sigma_d$ and area at the k-th binned region, respectively. We estimate the uncertainties of these expectation values of DGR with the ``realize'' method \citep{GORDON14}, and the uncertainties are $\sim$ 0.02 dex in the high metallicity region, $\sim$ 0.09 dex at $\metal\sim8.2$, and $\sim$ 0.6 dex in the lowest metallicity region, which are reasonably small. However, there is also intrinsic scatter of DGR in each radial region, which would be larger than the uncertainties. To estimate this intrinsic scatter of DGR per $M_{\rm gas}$ within one radial region, we calculate the distribution by summing up the PDFs of DGR from each bin in that radial region, weighted by their $M_{\rm gas}$. Next, we take the region between the 16th and 84th percentile of the distribution as the range of the intrinsic scatter. This intrinsic scatter is included in Figure \ref{fig: DGR_grad_models_merged}, along with the zero-point uncertainty in $\Sigma_{\rm gas}$.

\begin{figure}
\centering
\includegraphics[width=\columnwidth]{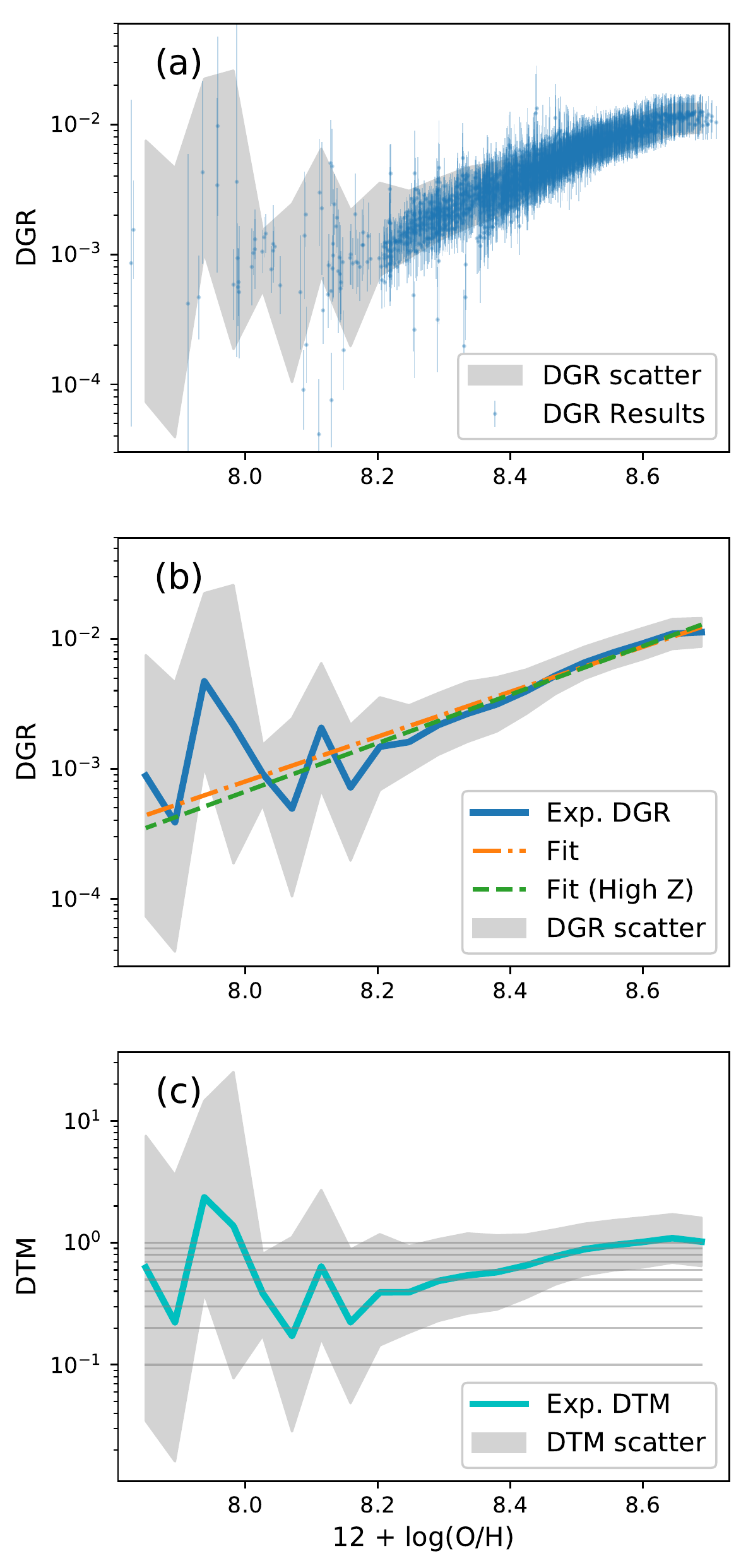}
\caption{Our DGR and DTM versus metallicity from the BE method results. (a) The DGR expectation values fitted by the BE model from each binned region are shown with error bars. Shaded region: The scatter of DGR. The definition is described in Figure \ref{fig: DGR_grad_models_merged}. (b) The DGR from the BE model with power-law (${\rm DGR} \propto Z^x$) fitting as listed in Table \ref{tab: D2M}. Blue: The expectation values calculated from the combined PDF (same for figures in \secref{sec: discussions}). Orange: The power-law result with whole data range. Green: Fitting with only $\metal > 8.2$, where we have a more concentrated data point distribution. (c) The DTM from the BE model. The DTM scatter includes DGR scatter, the $\metal$ uncertainty \citep{CROXALL16}, and $M_{\rm O}/M_{\rm Z}$ uncertainty (\secref{sec: MOMZ unc}). The horizontal lines are the DTM=0.1, 0.2, ......1.0 locations.\label{fig: DGR_grad_models_BE}}
\end{figure}
The distribution of our original data points is denser in the region with $\metal \ga 8.2$, where the original SNR is high. This is illustrated in Figure \ref{fig: DGR_grad_models_BE} (a) with the results from the BE model. Within this range, all models except SE have their DGR dropping by nearly 1 dex, which is around twice faster than the metallicity gradient. The SE has its DGR dropping by 1.5 dex. At $\metal < 8.0$, the scatter in PDF is large (generally with $\sigma \ga 1~\rm dex$), which makes determining a trend difficult. By treating metallicity as an independent variable, we fit our DGR versus metallicity with a linear equation ${\logt \rm DGR} = a\times(\metal) +b$ in both the full metallicity range and only $\metal \geq 8.2$ region. An example showing results from the BE model is shown in Figure \ref{fig: DGR_grad_models_BE} (b). The results are listed in Table \ref{tab: D2M}. All the fitting results indicate a $\logt{\rm DGR}$ variation steeper than $\metal$. The three methods with $\beta$ fixed over the whole spectral range, FB, WD, and PL, have fitted slopes closer to one.
\begin{table}
    \centering
    \caption{$\logt$DGR versus $\metal$ linear fitting results.\label{tab: D2M}}
    \begin{tabular}{ccccc}
        \hline
        \hline
        Model & \multicolumn{2}{c}{Full range} & \multicolumn{2}{c}{$\metal\geq 8.2$} \\
        & a & b & a & b \\
        \hline
        SE & $2.7\pm0.3$ & $-25.3\pm 2.1$ & $3.2\pm0.2$ & $-29.4\pm 1.8$ \\
        FB & $1.5\pm0.1$ & $-14.9\pm0.9$ & $1.5\pm0.1$ & $-15.3\pm 0.7$ \\
        BE & $1.7\pm0.1$ & $-16.9\pm 1.0$ & $1.9\pm 0.1$ & $-18.1\pm 0.7$ \\
        WD & $1.5\pm0.2$ & $-14.9\pm1.4$ & $1.3\pm0.1$ & $-13.1\pm0.5$ \\
        PL & $1.3\pm0.1$ & $-13.3\pm 0.9$ & $1.2\pm0.1$ & $-12.8\pm0.5$ \\
        \hline
    \end{tabular}
    \raggedright Note: Data are fitted with ${\logt \rm DGR} = a\times(\metal) +b$.
\end{table}

% Talk about upper limit
\subsubsection{Physical limitations to DGR}\label{sec: MOMZ unc}
Dust grains are built from metals. Thus, we can calculate the theoretical upper limit to the DGR by calculating the DGR for the case when all available metals are in dust. If the fitted DGR exceeds the calculated upper limit, we would consider the fitting result physically less plausible. To convert to total metallicity from oxygen abundance, we need to assume the ISM chemical composition. We calculate the mass ratio of oxygen to total metal from two literature of solar chemical composition: 1) \citet{LODDERS03}, which gives $M_{\rm O}/M_Z = 51\%$ where $M_Z$ is the mass of all metals. This is the composition used in \citet{JENKINS09}, which we will discuss in \secref{sec: J09}. 2) A later version in \citet{ASPLUND09}, which gives $M_{\rm O}/M_Z = 44.5\%$. The conversion from $\metal$ to metallicity is given by:
\begin{equation}\label{eq: DGR_limit}
    \frac{M_Z}{M_{\rm gas}} = \frac{M_Z}{M_{\rm O}}\frac{M_{\rm O}}{1.36M_{\rm H}} = \frac{\frac{m_{\rm O}}{m_{\rm H}}10^{\big(\metal\big)-12}}{\frac{M_{\rm O}}{M_Z} \times 1.36},
\end{equation}
where $m_{\rm O}$ and $m_{\rm H}$ are the atomic weights of oxygen and hydrogen. The solar $\metal$ adopted in this study is $8.69\pm0.05$ \citep{ASPLUND09}. This estimation of the DGR upper limit can be incorrect if the actual chemical composition deviates from this range. For example, \citet{CROXALL16} showed there is a trend that $\logt({\rm N/O})$ goes from $-0.4$ to $-1.4$ as radius increases in M101, which means we can overestimate the upper limit in the outer disk if other major elements have similar trends.

We overlay the DGR upper limit calculated between $M_{\rm O}/M_Z = 44.5\%$ and $51\%$ with our results in Figure \ref{fig: DGR_grad_models_merged}. We find that in the highest metallicity region, the DGR given by SE method is greater than the upper limit by a factor of 3, which is outside the 16-84 percentile of intrinsic scatter. This is unlikely being a result of $\alpha_{\rm CO}$ variation because we will need to have $\alpha_{\rm CO}\sim 9$ in the center of M101 to explain this apparent DGR. This $\alpha_{\rm CO}$ value is unlikely to be true with our knowledge of $\alpha_{\rm CO}$ in M101 \citep{SANDSTROM13} and metallicity-dependency of $\alpha_{\rm CO}$ \citep{BOLATTO13}. We thus consider the results from SE method less physically plausible.

We also notice that for all methods listed, there is a DGR spike in expectation value exceeding the upper limit near $\metal \sim 7.9$. Nevertheless, all the others still have their 16-84 percentile scatter falling under the DGR upper limit. Thus, we consider all methods except SE still reasonable under DGR upper limit test. Note that the scatter in the regions with $\metal < 8.2$ reach the order of 1 dex, which means the fit values are less reliable.

\subsection{Temperature profiles}\label{sec: T grad}
\begin{figure}
\centering
\includegraphics[width=\columnwidth]{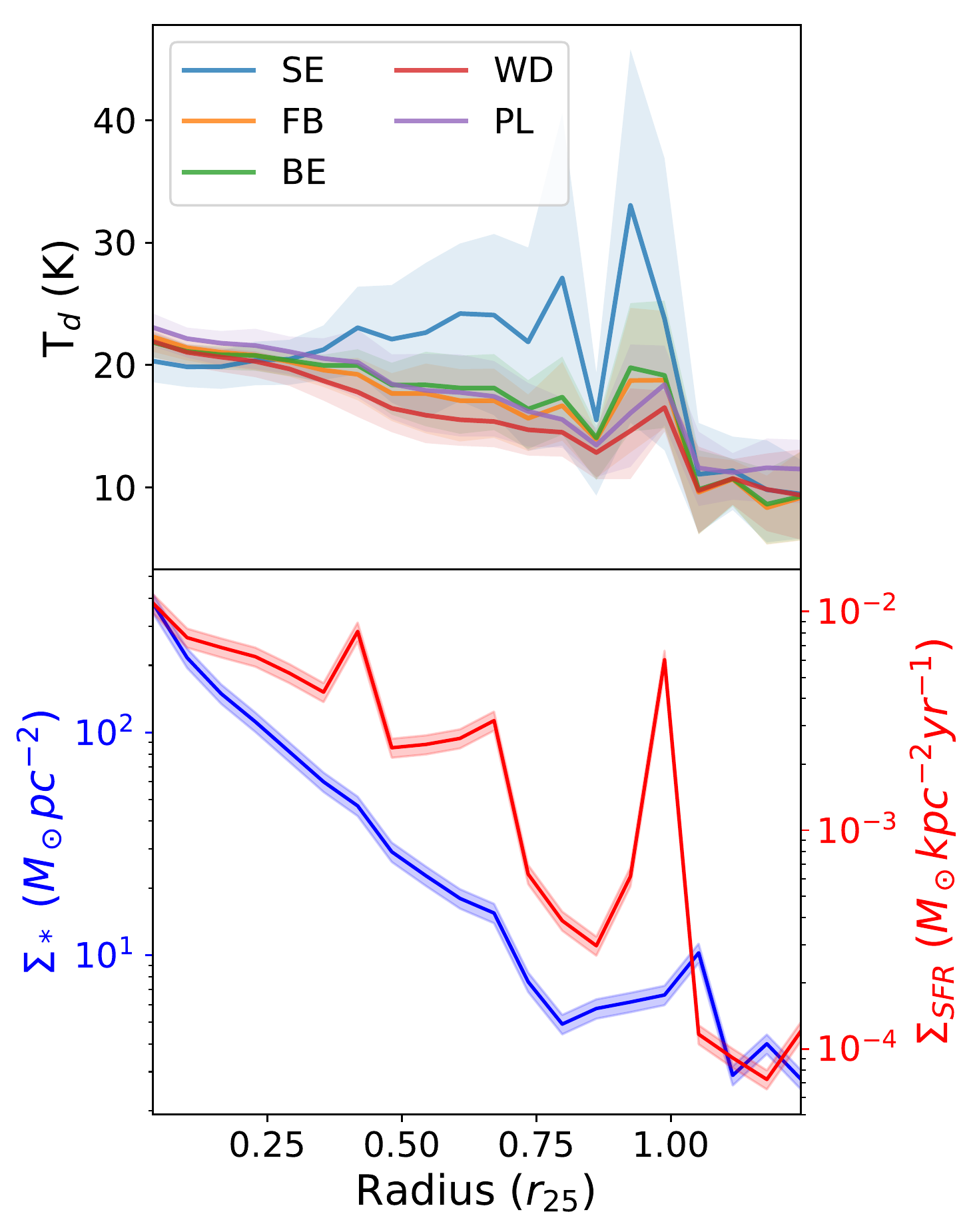}
\caption{Radial profiles of dust temperature, $\Sigma_{\rm SFR}$ and $\Sigma_\star$. All the profiles are plotted in gas-mass-weighted average. Top panel: temperature profiles from all fitting methods. 16-84 percentile scatter from the fitting is shown in shaded areas. Bottom panel: $\Sigma_{\rm SFR}$ and $\Sigma_\star$ profiles. See \secref{subsec: other} for data source and calculation. A 10\% uncertainty is plotted in shaded region, which is an uncertainty suggested in \citet{DALE09}.\label{fig: T_prof_new}}
\end{figure}
In the top panel of Figure \ref{fig: T_prof_new}, we plot the $M_{\rm gas}$-weighted dust temperature as a function of radius for each method. Within a small radial range, we assume that the DGR variation is small, thus the $M_{\rm gas}$-weighted dust temperature would be a representative $T_d$ in the corresponding radial region. For the PL method, temperature is not a directly fitted variable. Thus, we calculate the dust mass-weighted average $U$, and convert it to temperature according to \secref{sec: PL}.

The equilibrium dust temperature depends on the heating radiation field, which should be related to a combination of $\Sigma_\star$ and $\Sigma_{\rm SFR}$ here, shown in bottom panel of Figure \ref{fig: T_prof_new}. By comparing to the radial trend of heating sources, the one model that stands out is SE: it has a temperature profile rising from the galaxy center to $0.8R_{25}$. It is possible to change the relationship between heating sources and dust temperature if the geometry and/or the opacity of the ISM changes with radius. However, with both heating source tracers having intensity decreasing by more than one dex within $0.8R_{25}$, we expect that a decreasing $T_d$ with radius to be the dominant trend. Thus, we also reach the conclusion as previous section that results from SE method are less physically plausible.

\subsection{Residual distributions}
\begin{figure*}
\centering
\includegraphics[width=\textwidth]{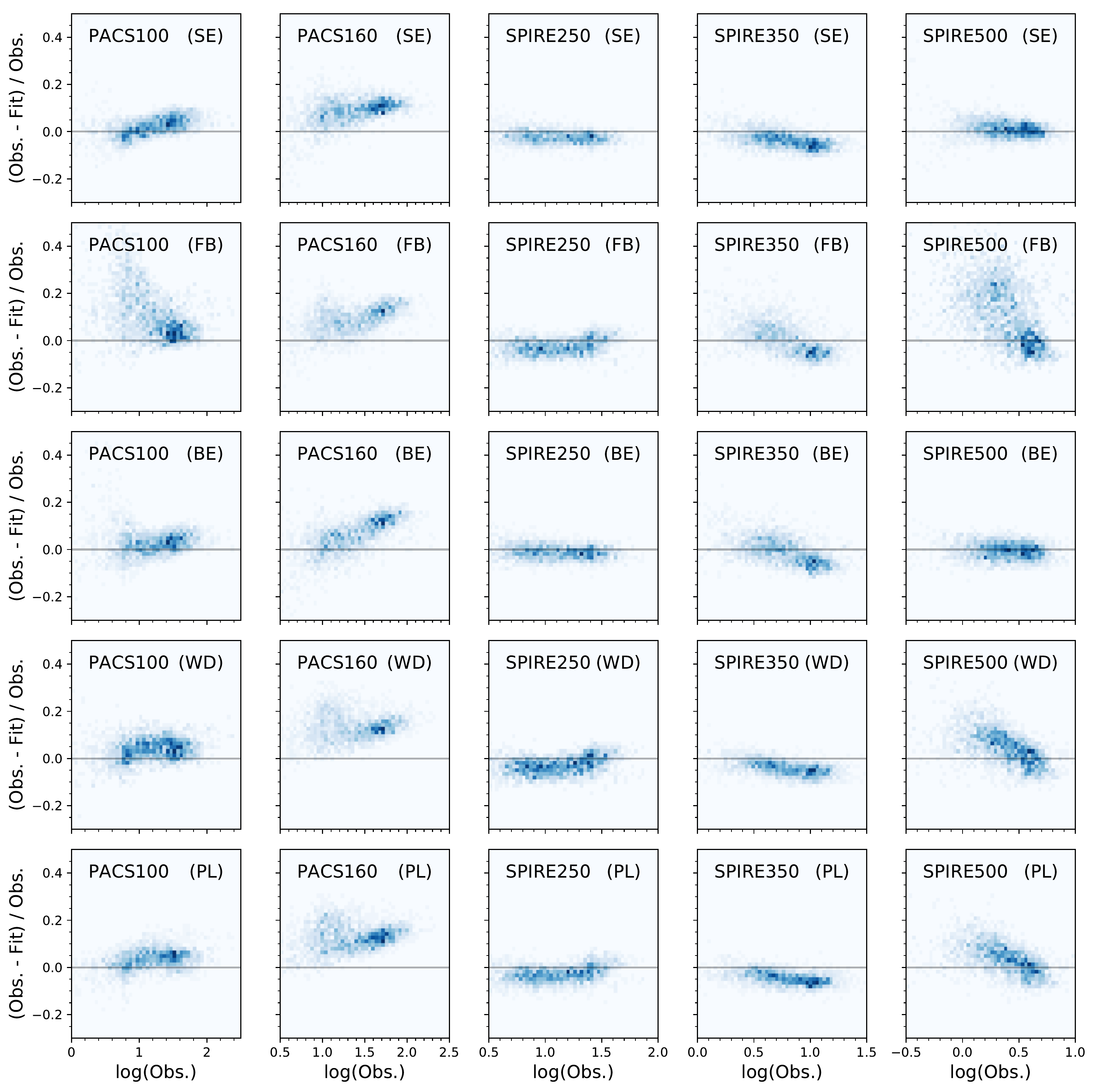}
\caption{2-dimensional histograms of relative residual versus observed SED at each band. The x-axes have unit in $\rm MJy~sr^{-1}$. The zero relative residual line is marked in gray.\label{fig: Residual_merged}}
\end{figure*}
The residual distribution is one of the most straightforward ways to check the goodness of fit. For each method, we plotted the 2-dimensional histogram of relative surface brightness residuals in Figure \ref{fig: Residual_merged}. We expect that a good fit will give a residual distribution that is symmetric about zero (The gray lines in all panels in Figure \ref{fig: Residual_merged}) and has no trend with the measured surface brightness. An example of well-behaved residual distribution can be seen for the BE model at the SPIRE 250 band. Otherwise, there may be a underlying systematic effect which tells us that the model is flawed or an additional free parameter is needed.

There are two features occurring for all MBB methods: 1) At the high intensity end, all of our methods underestimate PACS160. 2) In general, the relative residuals are smaller at the low intensity end (see more discussion in \secref{sec: compare-chi2}). The SE method gives the most compact residual distributions. This means that letting both $T_d$ and $\beta$ free provides the highest flexibility to fit the SED among all models here. However, we should bear in mind that the SE model yields DGR and temperature gradients distinct from the other models and that we consider these results less physically plausible, as previously shown in \secref{sec: max DGR} and \ref{sec: T grad}. The FB method yields the residual distribution least consistent with random scatter about the model. It shows the least compact residual distribution with long tails in positive residuals, especially in PACS100, SPIRE350 and SPIRE500. These positive residuals mainly come from low intensity regions. These indicate the need for $\beta$ to change between high and low intensity regions. Among the remaining methods, both the WD and PL improve the residuals the short wavelengths covered by PACS100. This reflects the expected presence of warm, possibly out-of-equilibrium dust at these short wavelengths. The BE method has the second most compact residual distribution, and shows a better fit to the long wavelength bands that are crucial to accurately trace $\Sigma_d$.

\subsection{The reduced chi-square values\label{sec: compare-chi2}}
\begin{figure}
\centering
\includegraphics[width=\columnwidth]{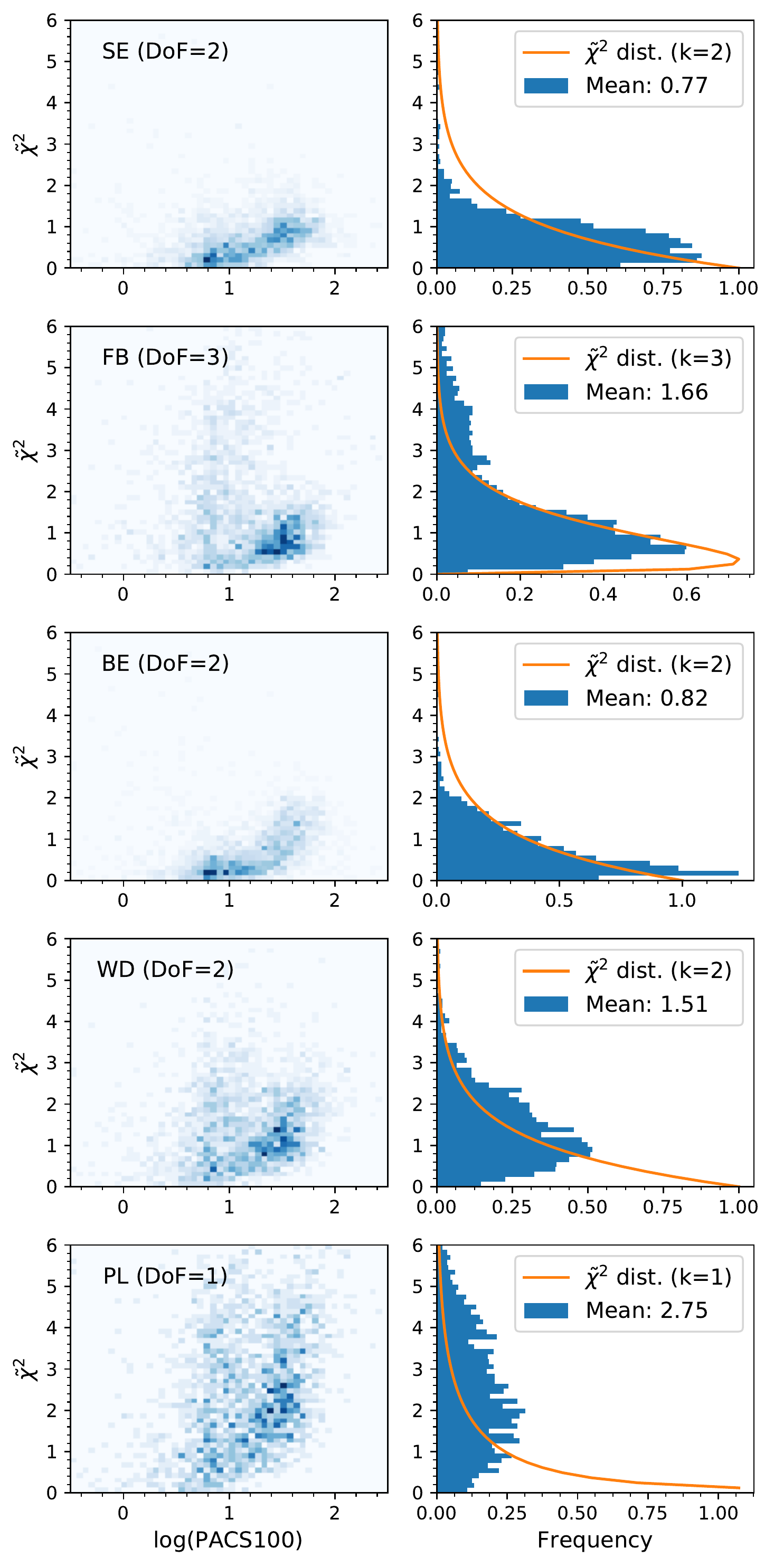}
\caption{The $\tilde{\chi}^2$ distributions for all fitting methods. Left: 2-dimensional histograms of $\tilde{\chi}^2$ with PACS100. The x-axes have unit in $\rm MJy~sr^{-1}$. Note that the 2-dimensional histograms of $\tilde{\chi}^2$ with all five bands demonstrate similar information, thus we only plot the ones from PACS100. Right: the horizontal histograms of $\tilde{\chi}^2$. The orange lines show the expected distribution according to DoF.\label{fig: residuals_chi2}}
\end{figure}
The reduced chi-square value is defined as $\tilde{\chi}^2 \equiv \chi^2/(n - m)$, where $n$ is the number of observations (which is $5$ in our study) and $m$ is the number of fitting parameters (3 for SE, 2 for FB, 3 for BE, 3 for WD and 4 for PL). This value takes both uncertainties in the observations and the degrees of freedom (DoF) of the models into account. The $\tilde{\chi}^2$ value gives the information of how good the fitting is and how much an extra fitting parameter improves the fitting quality. We plot the $\tilde{\chi}^2$ distribution versus observation in the left panels in Figure \ref{fig: residuals_chi2}. As we have seen in residual maps, the FB and WD methods have long tails in the low luminosity region. FB and WD methods have $\tilde{\chi}^2 \geq 1$ in the high luminosity region, mainly due to the residuals in long wavelength, where the corresponding uncertainties are much smaller. The PL method has relatively large $\tilde{\chi}^2$ everywhere, which means the extra DoF does not offer an improvement in the quality of the fitting. Note that this result does not imply the physical correctness of single temperature over ISRF distribution, but indicates that the DoF from ISRF distribution is less effective in improving the quality of FIR SED fitting.

All the methods have a gradually rising $\tilde{\chi}^2$ toward the high luminosity region. By calculating the contribution to $\tilde{\chi}^2$ from each band, the most important contributor to this phenomenon is the PACS160 band. There is in general a $\sim$20\% systematic underestimation by the model fits in PACS160 in the center of M101. One possible explanation is that the contribution from \textsc{[Cii]} 158 \micron\ line is integrated into the PACS160 SED, which makes the PACS160 SED brighter than what is predicted by dust emission models. This effect is shown to be minor by \citet{GALAMETZ14}, where the authors demonstrated that \textsc{[Cii]} contributes only around 0.4\% to integrated 160~\micron\ emission. Another possible explanation is an unknown systematic uncertainty in PACS160. Previous work by \citep{ANIANO12} found that PACS160 was $\sim$20\% higher than {\em Spitzer} MIPS160 measurements in the bright regions of some nearby galaxies.

We also examine the histograms of $\tilde{\chi}^2$ (Figure \ref{fig: residuals_chi2} right panels) with two features: 1) The mean value, which is expected to be one. 2) The shape of the histogram, which should resemble the $\chi^2$-distribution with $k$ DoF\footnote{We normalized the $\chi^2$-distribution to a mean value of one, i.e., $k\times f(k\tilde{\chi}^2, k)$.}. The SE method has mean $\tilde{\chi}^2$ of 0.77. The histogram is more compact than a $\chi^2$-distribution with $k=2$. Both indicate that we might be overestimating the uncertainties in the SE method. FB and WD have mean values of 1.5 and 1.64, respectively, and flatter histograms than expected. BE has a mean value of 0.97 and a distribution resembling what we expected. PL has a mean value of 3.16, which means the extra parameters in the PL model do not help it making a more precise fit corresponding to its DoF.

\subsection{Summary of model comparison}
Among the MBB variants we have tested, we consider the SE method physically less plausible because the resulting temperature and DGR gradient do not match our physically-motivated expectations. The DGR results from the other four MBB variants are consistent with each other in regions with $\metal \leq 8.5$, as illustrated in Figure \ref{fig: DGR_grad_models_merged}. This implies that the dust masses measure from the MBB fitting is mostly insensitive to the specific choices about the radiation field distribution. According to the residual distribution and $\tilde{\chi}^2$ values, the BE model gives the statistical best fit, which means that the most important first-order correction to the basic MBB is to allow $\beta$ vary in the long wavelength region. We will consider BE as the preferred model based on these tests.

\section{Discussion}\label{sec: discussions}
\subsection{Is DTM Constant in M101?}\label{sec: dicuss_variable_DTM}
% Our fitting result: variable DTM
All of our models indicate that DGR falls off steeper than metallicity, showing a variable DTM ratio. Our preferred model (BE) has ${\rm DGR} \propto Z^{1.7}$, which is equivalent to DTM changing from 0.25 at $\metal\sim$7.8 to 1 above $\metal\sim$8.5. Models with $\beta$ fixed have smaller power-law indices, specifically the FB and WD models show ${\rm DGR} \propto Z^{1.4}$, and PL model shows ${\rm DGR} \propto Z^{1.2}$. Even if we only consider region with $\metal \geq 8.2$, where the majority of our data points reside, we still obtain a DGR trend steeper than metallicity gradient. These results are based on direct-$T_e$ method metallicity measurements \citep{CROXALL16} with uncertainties in $\metal$ around $0.04-0.08$ dex.

In order to understand what aspects of the dust life cycle could result in a variable DTM, we look for mechanisms that affect dust mass and metals in the ISM with different rates. The five most important mechanisms of this kind are: 1) Accretion of metals in the ISM onto existing dust grains, which raises DTM. 2) ISM enrichment from stellar sources (e.g. AGB stars, SNe), which have DTM characteristic of the particular stellar source instead of DTM in the current ISM. 3) Dust destruction by SNe, which lowers DTM. 4) Infall of circumgalactic medium (CGM) into the galaxy, which dilutes the ISM DTM with the lower DTM in the CGM \citep{DWEK98, HIRASHITA99, ZHUKOVSKA16}. 5) Outflows of dust and metals into CGM, which increases the ISM DTM because the outflow is less dusty than the ISM \citep{LISENFELD98}.

Among these mechanisms, ISM accretion has a rate that increases with ISM density, especially in cold clouds \citep{DWEK98, ASANO13}. Observationally, ISM density can be roughly traced by the mass fraction of molecular hydrogen ($\rm f_{H_2}$)\footnote{Without knowing the three-dimensional ISM geometry, $\rm f_{H_2}$ would be a better indicator of ISM density than $\Sigma_{\rm gas}$.}. The rate of enrichment from stellar sources should follow the stellar mass surface density ($\Sigma_\star$) modulo stellar age effects. The effects of production and destruction of dust by SNe should track both the massive star formation rate ($\Sigma_{\rm SFR}$) and the older stellar populations ($\Sigma_\star$).
\begin{table}
    \centering
    \caption{Correlation between $\logt$DTM and physical quantities $\logt\rm f_{H_2}$, $\logt\Sigma_\star$ and $\logt\Sigma_{\rm SFR}$.\label{tab: Residual trend}}
    \begin{tabular}{ccccc}
        \hline
        \hline
        Quantity & \multicolumn{2}{c}{Direct} & \multicolumn{2}{c}{Residual} \\
        & $\rho_S$ & $p$-value\footnote{$p$-value is the probability that we get a $\rho_S$ greater or equal to the calculated value from the given data when null hypothesis is true. In other words, $p$-value goes from 0 to 1, and a smaller $p$-value implies a more significant correlation.} & $\rho_S$ & $p$-value \\
        \hline
        $\logt\rm f_{H_2}$ & 0.80 & $\ll 1$ & 0.26 & $\ll 1$ \\
        $\logt\Sigma_\star$ & 0.72 & $\ll 1$ & -0.05 & 0.12 \\
        $\logt\Sigma_{\rm SFR}$ & 0.22 & $\ll 1$ & -0.08 & 0.007 \\
        \hline
    \end{tabular}
\end{table}

\begin{figure*}
\centering
\includegraphics[width=\textwidth]{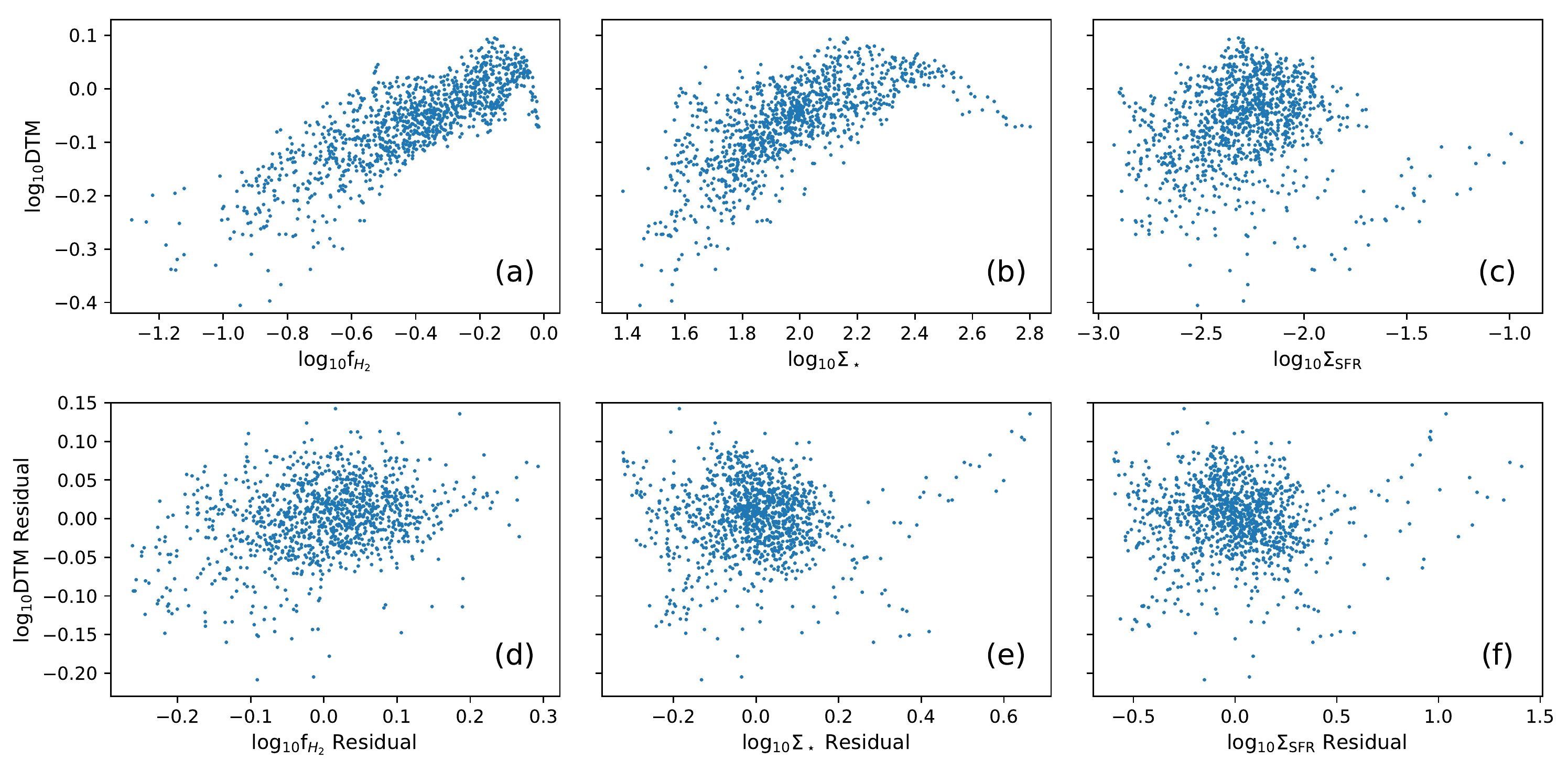}
\caption{Relation between DTM and the three physical quantities: $\rm f_{H_2}$ (a, d), $\Sigma_\star$ (b, e) and $\Sigma_{\rm SFR}$ (c, f). (a, b, c): Relations in the raw data. (d, e, f): Relations after removing the radial trends in all four quantities: $\logt$DTM, $\logt\rm f_{H_2}$, $\logt\Sigma_\star$ and $\logt\Sigma_{\rm SFR}$. The radial trend removal is done by first fitting the quantities versus radius with linear regression, and then subtracting the regression results from the original data. The discussion of radial trend removal is described in \secref{sec: dicuss_variable_DTM}. The mean uncertainty in $\Sigma_d$ is 0.1 dex. $\Sigma_\star$ has unit in $M_{\sun}~\rm pc^{-2}$ and $\Sigma_{\rm SFR}$ has unit in $M_{\sun}~\rm kpc^{-2}~yr^{-1}$.\label{fig: residual trend}}
\end{figure*}
To test these potential correlations of DTM with environmental characteristics, we calculate the Spearman's rank correlation coefficient ($\rho_S$) and $p$-value between $\logt$DTM and these three quantities. Note that we only include the region with $\rm f_{H_2}\geq 5\%$ for all four quantities, namely DTM, $\rm f_{H_2}$, $\Sigma_\star$ and $\Sigma_{\rm SFR}$, due to the detection limit of HERACLES. $\logt$DTM correlates strongly and significantly with both $\logt \rm f_{H_2}$ and $\logt \Sigma_\star$, while it shows a weaker but significant correlation with $\logt\Sigma_{\rm SFR}$. This is shown in the ``direct'' columns in Table \ref{tab: Residual trend} and top panels in Figure \ref{fig: residual trend}.

While there are significant correlations between DTM and these environmental characteristics, all the quantities here (DTM, $\rm f_{H_2}$, $\Sigma_\star$ and $\Sigma_{\rm SFR}$) to first order have major trends that vary with radius. $\rm f_{H_2}$, $\Sigma_\star$ and $\Sigma_{\rm SFR}$ all have $\rho_S$ with radius greater than the $\rho_S$ with $\logt$DTM. $\logt$DTM also has a higher $\rho_S$ with radius than with other quantities. The results of calculating the $\rho_S$ and $p$-value directly will therefore be dominated by this major radial trend. In order to investigate what drives the DTM variation, we need to remove these dominant radial trends. This removal is done by first fitting $\logt$DTM, $\logt \rm f_{H_2}$, $\logt\Sigma_\star$ and $\logt\Sigma_{\rm SFR}$ versus radius with linear regression, and then subtracting the regression results from the original data points to get the residuals. The correlations between $\logt$DTM and $\logt \rm f_{H_2}$, $\logt\Sigma_\star$ and $\logt\Sigma_{\rm SFR}$ after radial trend removal are shown in the bottom panels in Figure \ref{fig: residual trend} and the ``Residual'' columns in Table \ref{tab: Residual trend}.

The resulting $\rho_S$ between residual $\logt$DTM and residual $\logt\rm f_{H_2}$ is 0.26, with a $p$-value $\ll 1$. This indicates that the correlation between them is weak compared to the scatter in the data but significant. The null hypothesis, that the two variables (residual DTM and $\rm f_{H_2}$) are unrelated, is extremely unlikely to be true. Residual $\logt\Sigma_\star$ and residual $\logt\Sigma_{\rm SFR}$, on the other hand, have their $\rho_S$ drop relative to the direct correlation and the residual $\rho_S$ of them show extremely weak correlations, and thus considered negligible.

Based on this calculation, we suggest that ISM density may be the most important environmental factor that affects DTM in M101. This would explain the correlation between variations of DTM at a fixed radius and variations in $\rm f_{H_2}$. The stellar sources, traced by $\Sigma_\star$ and $\Sigma_{\rm SFR}$, do not correlate significantly with the variations of DTM at a fixed radius.

\subsubsection{Variable emissivity coefficient}\label{subsec: discuss_emissivity}
% Alternative explanation.
Although we have thus far interpreted our results as changes in DTM, an alternative possibility is that $\kappa_{160}$ varies with environment instead. As discussed in \secref{Sec: Model} all of our MBB variants are subject to the degeneracy between $\Sigma_d$ and $\kappa_{160}$. The way we deal with it is by calibrating $\kappa_{160}$ with the MW diffuse ISM SED (\secref{Sec: calibration}) and assuming all the variation in temperature-corrected SED amplitude is due to $\Sigma_d$ only. However, this assumption might fail if we observe environments that differ from the high-latitude MW diffuse ISM we used for calibration and if $\kappa_{160}$ varies with local environment. In general, our DGR($Z$) does not follow the DGR($Z$) calculated from ${\rm F_\star}=0.36$, which have been used for our calibration. This leaves the possibility that the changes we see in DTM are still degenerate with the changes in $\kappa_{160}$.

$\kappa_{160}$ can be a function of dust size, temperature, and composition, which may change as gas transitions from diffuse to dense phases. The calculations in \citet{OSSENKOPF94, KOHLER11} show an enhanced dust emissivity due to coagulation of dust particles in dense ISM regions. This phenomenon is also observed by \citet{PLANCK14} and \citet{PLANCK15} in the MW, where the authors show an increase in total opacity with increasing ISM density and decreasing $T_d$. However, we note that both \citet{PLANCK14} and \citet{PLANCK15} \textit{assumed} a constant DGR, and explained their observations with a change in the composition and structure of the dust particles.

\begin{figure}
\centering
\includegraphics[width=\columnwidth]{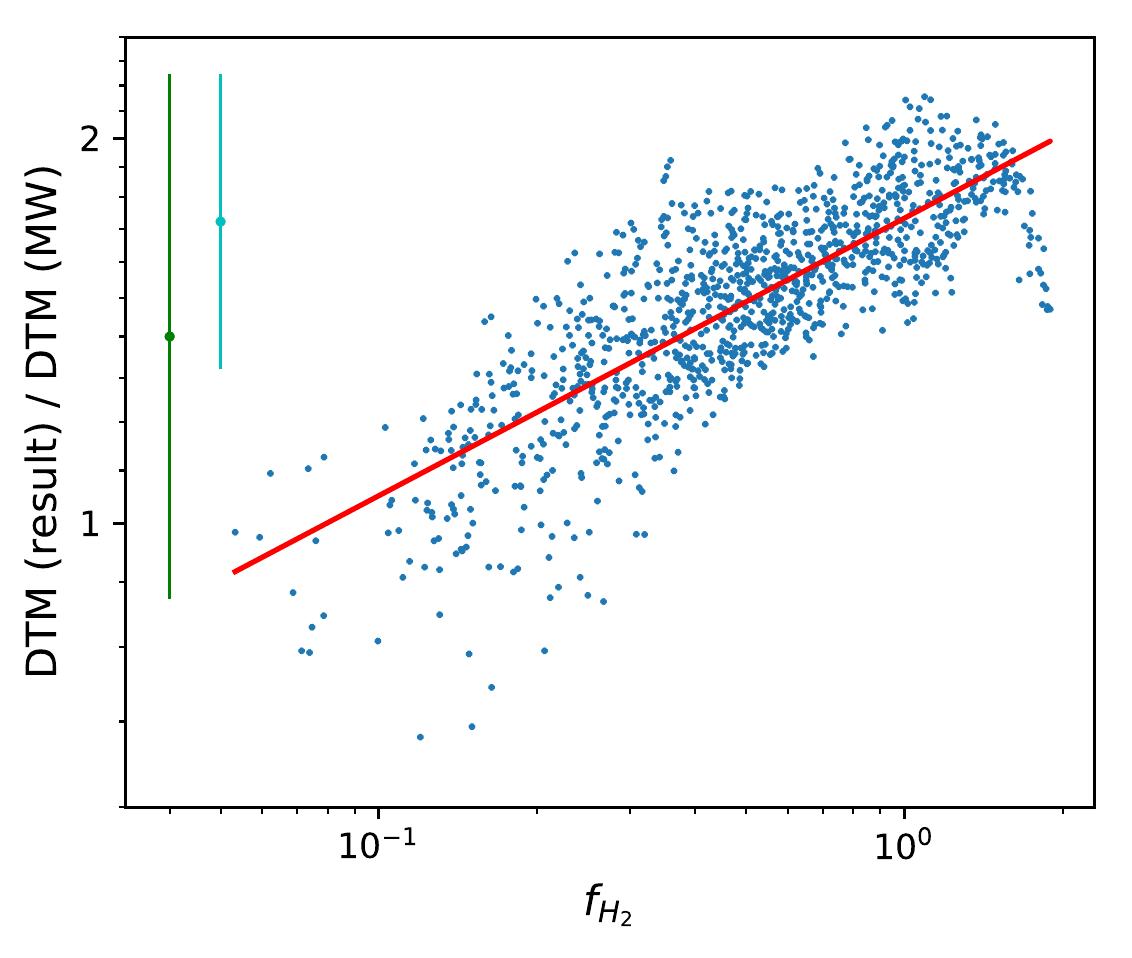}
\caption{Our DTM normalized by the MW DTM \citep{DRAINE11} plotted as a function of H$_2$ mass fraction ($f_{\rm H_2}$). The original distribution is shown in blue. A representative error bar in cyan, which only include the uncertainties in DGR, is shown at top-left. Another error bar including extra uncertainty in $\metal$, which is considered systematic, is shown in green at top-left. The linear regression of $\logt \rm DTM/DTM_{MW}$-to-$\logt f_{\rm H_2}$ is shown in red. Note that this plot only includes data with $f_{\rm H_2} \ga 5\%$ ($\metal \ga 8.4$), and that the y-axis is in log scale.\label{fig: rDGR_vs_metal_BE}}
\end{figure}

We will focus on the dense regions in M101 for discussing emissivity variation with coagulation, where coagulation is more likely to happen. 
We use the constant DTM in MW \citep{DRAINE11} as our reference true DTM and calculate how our DTM deviates from the reference as a function of ISM density, traced by $\rm f_{\rm H_2}$, plotted Figure \ref{fig: rDGR_vs_metal_BE}. Note that the figure only includes the region with significant detection from HERACLES ($f_{\rm H_2} \ga 5\%$, or $\metal \ga 8.4$), not the full range of our DGR-to-metallicity figures.

We calculate the Pearson's correlation coefficient of all four combinations of log/linear $\frac{DTM}{DTM_{MW}}$-to-$ f_{\rm H_2}$ relation, i.e., $\rm \frac{DTM}{DTM_{MW}}$-to-$f_{\rm H_2}$, $\rm \frac{DTM}{DTM_{MW}}$-to-$\logt f_{\rm H_2}$, $\logt \rm \frac{DTM}{DTM_{MW}}$-to-$f_{\rm H_2}$, and $\logt \rm \frac{DTM}{DTM_{MW}}$-to-$\logt f_{\rm H_2}$. The result shows 0.712, 0.790, 0.694, and 0.795, respectively. Thus we continue our analysis with $\logt \rm \frac{DTM}{DTM_{MW}}$-to-$\logt f_{\rm H_2}$ relation. By fitting $\logt \rm \frac{DTM}{DTM_{MW}}$ to $\logt f_{\rm H_2}$, our $\rm \frac{DTM}{DTM_{MW}}$ varies from 0.9 to 2.0 in this region. If we attribute this change to the increase in emissivity, then $\kappa_{160}$ will go from 19 to $41~\rm cm^2~g^{-1}$ in this region, with a relation of $\kappa_{160} \propto f_{\rm H_2}^{0.2}$. This is comparable to the emissivity changes inferred by \citet{PLANCK14} using similar reasoning in MW clouds and well within the range allowed by theoretical grain coagulation models \citep{OSSENKOPF94, KOHLER11}.

\subsubsection{Variable conversion factor}\label{sec: alpha_CO discussion}
Another potential explanation of the change in DGR (and thereby DTM) is that the conversion factor $\alpha_{\rm CO}$ is not a constant, therefore, we could be wrong in estimating $\Sigma_{\rm H_2}$. There are two major observed trends in $\alpha_{\rm CO}$ \citep{BOLATTO13}. The first trend is a metallicity dependent $\alpha_{\rm CO}$. In the model derived in \citet{WOLFIRE10}, among others, $\alpha_{\rm CO}$ increases as metallicity decreases, which means we could be overestimating DGR in the outer part of M101. Recovering this overestimation would increase the variation in DTM and make the observed trends stronger. Moreover, since $\rm f_{H_2}$ traced by a fixed $\alpha_{\rm CO}$ drops steeply with increasing radius in M101, any modification from metallicity dependent $\alpha_{\rm CO}$ that can affect DGR in the disk must posit a large and almost totally invisible reservoir of CO-dark molecular gas. It is suggested by \citet{BOLATTO13} to use a constant $\alpha_{\rm CO}$ in regions with $\metal\ge 0.5Z_\sun$. When we test the total gas mass from a constant $\alpha_{\rm CO}$ against the one calculated with \citet{WOLFIRE10} metallicity-dependent $\alpha_{\rm CO}$, the difference between them is at most 0.12 dex. This small change is due to the fact that in the radial region of M101 where H$_2$ makes a substantial contribution to the total gas mass, the metallicity is greater than $\metal=8.4$, where $\alpha_{\rm CO}$ only changes by a small amount. Considering the unknown uncertainties caused by the constant DTM assumption in the metallicity-dependent model \citep{BOLATTO13}, we decide to only present the results with a fixed $\alpha_{\rm CO}$.

The second trend is the decrease of $\alpha_{\rm CO}$ in the very center of some nearby galaxies, shown by \citet{SANDSTROM13}. It is worth noting that the \citet{SANDSTROM13} analysis assumed DGR was locally independent of $\rm f_{H2}$ to simultaneously solve for $\alpha_{\rm CO}$ and DGR in their solution pixels. Over most of M101, however, the average $\alpha_{\rm CO}$ they find is similar to the standard MW conversion factor, so using the \citet{SANDSTROM13} values or making the standard assumption of a MW $\alpha_{\rm CO}$ will not greatly impact our results. \citet{SANDSTROM13} found that M101 has one of the largest observed central decreases in $\alpha_{\rm CO}$, showing $\alpha_{\rm CO}=0.35^{+0.21}_{-0.13}$ in the central solution pixel, which is far lower than the galaxy-average value. Adopting the galaxy average value of $\alpha_{\rm CO}$ therefore causes us to overestimate the amount of gas in the center and subsequently underestimate the DGR and DTM. As shown in Figure \ref{fig: DGR_grad_models_BE} (c), we do observe a decrease in the DGR and DTM in the central $\sim$kpc of M101, which is likely the result of an incorrect conversion factor assumption there. However, since the affected region is small compared to our full M101 maps, we can neglect this effect in the DTM discussion.

Beyond radial trends that alter $\alpha_{\rm CO}$ relative to what we have assumed, it is also possible that $\alpha_{\rm CO}$ varies from cloud-to-cloud at a fixed radius. If we overestimate $\alpha_{\rm CO}$ for a cloud, the DTM would be underestimated and $\rm f_{H_2}$ would be overestimated. If we underestimate $\alpha_{\rm CO}$, we would underestimate $\rm f_{H_2}$ and overestimate DTM. Both overestimation and underestimation work in the opposite sense of the correlation we observe in the residual DTM and $\rm f_{H_2}$ and, if corrected for, would therefore strengthen our conclusions. Thus, the positive correlation between DTM and $\rm f_{H_2}$ we calculate previously is not a result of $\alpha_{\rm CO}$ variation.

\subsubsection{Summary of DTM Measurements}
To summarize, we can explain our fitting results from all our MBB variants except the SE model with a variable DTM, where ${\rm DGR} \propto Z^{1.7}$ in the BE model. The maximum DGR is still within the available total metal abundance limits. By comparing the correlation between DTM and physical quantities $\rm f_{H_2}$, $\Sigma_\star$ and $\Sigma_{SFR}$, we conclude that the strongest environmental correlation of DTM is with $\rm f_{H_2}$, which we take to be a reasonable observational indicator of ISM density and thus a tracer for accretion process. We see no clear trends that indicate correlations of DTM with stellar sources or massive star formation.

On the other hand, we could also explain the DTM results with enhanced dust emissivity in dense regions due to coagulation. The increase in $\kappa_{160}$ is at most twice of the originally calibrated value, which is within the findings in \citet{PLANCK14}. A non-extreme metallicity-dependent $\alpha_{\rm CO}$ does not affect our DGR trend much due to the low $\rm f_{H_2}$ in most regions, however, the change of $\alpha_{\rm CO}$ in the center is related to our observed decrease of DGR in the central kpc. Variability of $\alpha_{\rm CO}$ from cloud to cloud at fixed radius would lead to a negative correlation between residual DTM and residual $\rm f_{H_2}$, which is opposite what we observe.

Both explanations of variable DTM and variable emissivity are within the physically plausible range, thus we cannot definitively conclude if the variations we see are mainly due to changes in DGR or changes in the emissivity. However, given the observation that elemental depletions in the Milky Way are a function of ISM density and $\rm f_{H_2}$ \citep[][see further discussion below]{JENKINS09}, which is equivalent to a variable DTM, we argue that attributing all variation to emissivity is unlikely. To break the degeneracy between emissivity and $\Sigma_d$, one future path is to calculate emissivity from dust models according to physical properties of local ISM. Another is to build an observational database of $\Sigma_d$-to-SED, with known metallicity and ISM density, for future calibration. 
Another powerful test available in the near future will be to measure the properties of the UV/optical extinction curve, like $\rm R_V$, as a tracer for coagulation and processes that can change the IR emissivity in the Local Group, and correlate this extinction curve tracer with quantities observable outside the Local Group.

\subsection{Comparison with previous DTM studies}
\begin{figure}
\centering
\includegraphics[width=\columnwidth]{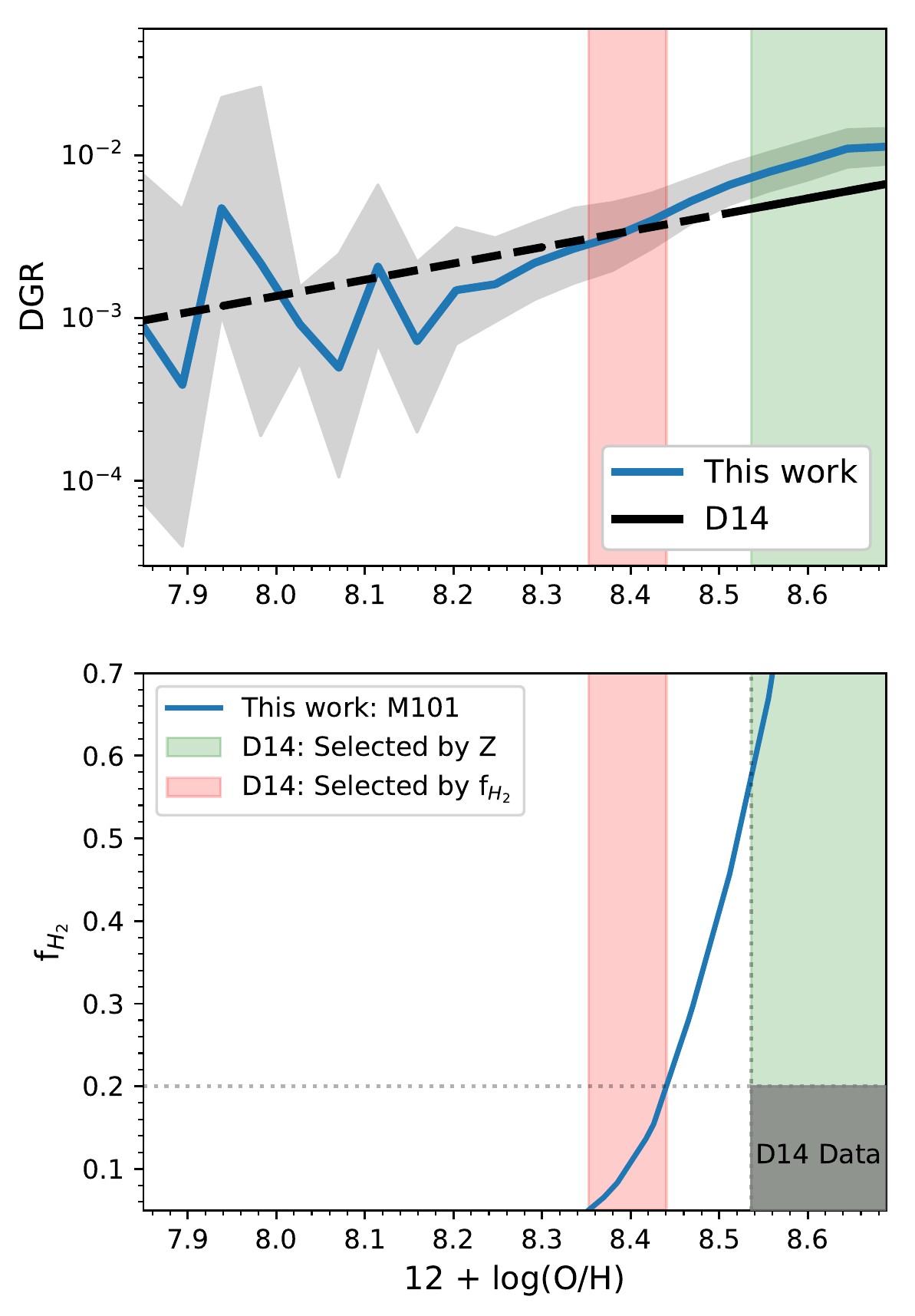}
\caption{Top: We compare our DGR($Z$) with that from M31 measured by \citet{DRAINE14}. The solid line is where \citet{DRAINE14} presents their DGR fitting in M31 with observed metallicity, and the dashed line is extrapolation of their linear DGR($Z$). Within this metallicity region, our M101 results suggest a DTM 2 times higher than M31. However, if we instead select the range of radii where the M31 $\rm f_{H_2}$ matches what we see in M101 (red region), we find a much better agreement between our observed DTM and extrapolation of \citet{DRAINE14}. Bottom: Demonstration of how we select the green and red zones. Grey zone: \citet{DRAINE14} $\rm f_{H_2}$ range corresponding to the presented $\metal$ range. Blue: $\rm f_{H_2}$-metallicity relation in M101. Green zone: Region with the same metallicity as \citet{DRAINE14} data range. Red zone: Region where M101 $\rm f_{H_2}$ corresponds to \citet{DRAINE14} $\rm f_{H_2}$.\label{fig: DTM_D14}}
\end{figure}
In Figure \ref{fig: DTM_D14}, we plot our results compared to the linear DGR($Z$) relation discussed in \citet{DRAINE14}. \citet{DRAINE14} show that the M31 DTM matches very well with the DTM predicted from depletions along the line of sight to $\zeta$Oph in the MW \citep[${\rm F_\star}=1$ line of sight in][]{JENKINS09}. In the corresponding metallicity range, our DGR is larger than the one in \citet{DRAINE14}. This is illustrated in Figure \ref{fig: DTM_D14} green zone. The derived $\kappa_{160}$ value in \citet{DRAINE14} is 12.51, which is around 0.75 times of our $\kappa_{160}$ value. Thus, the DGR discrepancy at high metallicity end is not a result of our choice of $\kappa_{160}$. Moreover, \citet{DALCANTON15, PLANCK16} indicates that the \citet{DRAINE07} model might overestimate $\Sigma_d$ by $\sim 2$ times, which also makes the difference larger. Thus, The difference between \citet{DRAINE14} and our results in high metallicity region is not due to parameter selection, but due to physical differences between M101 and M31, or differences in the modeling.

Instead of comparing region with the same metallicity, we can also compare the DTM between regions in M31 and M101 with similar ISM density, traced by $\rm f_{H_2}$ here. According to \citet{NIETEN06}, the region in M31 where \citet{DRAINE14} gives the direct metallicity measurements has $\rm f_{H_2}$ below 0.2, marked by the horizontal dashed line in Figure \ref{fig: DTM_D14}. This $\rm f_{H_2}=0.2$ upper limit meets our M101 data at $\metal=8.44$, indicated at where the horizontal dashed line meets the blue curve in Figure \ref{fig: DTM_D14}. We pick the region between $\metal=8.44$ and where we have minimum $\rm f_{H_2}$, shown in red in Figure \ref{fig: DTM_D14}, as the region that has similar ISM density with M31 data in \citet{DRAINE14}. Within this region, our DTM is consistent with the extrapolation of \citet{DRAINE14} DTM. This suggests that the difference in DTM between our results and \citet{DRAINE14} may be a consequence of M101 having a higher $\rm f_{H_2}$ and therefore enhanced depletion (e.g. larger DTM) at the metallicity of M31.

\begin{figure}
\centering
\includegraphics[width=\columnwidth]{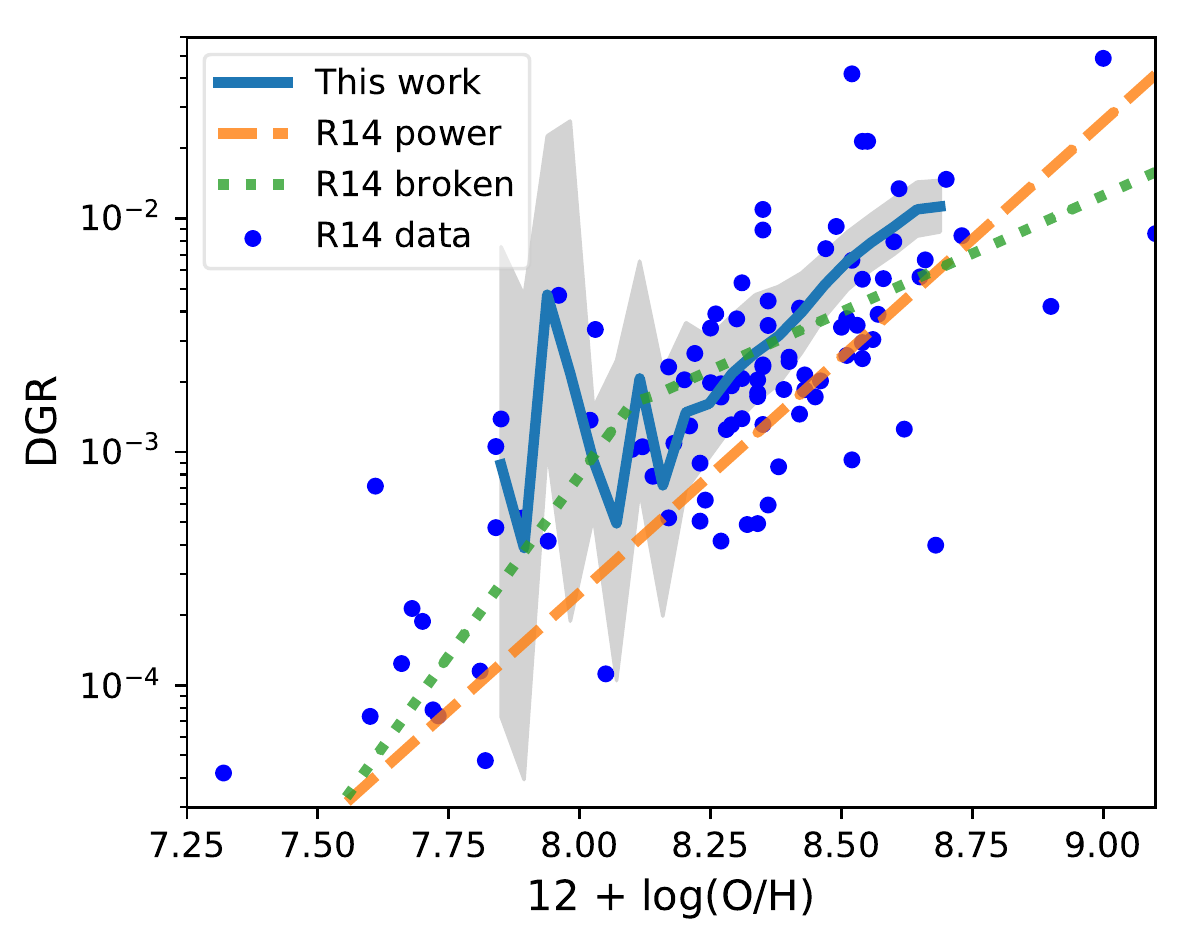}
\caption{Our DGR versus metallicity with \citet{REMY-RUYER14} results (data points in blue). The power law (orange dahsed line) and broken power law (green dotted line) fitting are quoted with MW conversion factors.\label{fig: DTM_R14}}
\end{figure}
\citet{REMY-RUYER14} has compiled integrated DGR($Z$) for a large set of galaxies observed by \textit{Herschel}. In Figure \ref{fig: DTM_R14} we compare our measured DGR($Z$) with theirs. At the high metallicity end, our slope is shallower than their power law fitting, but within 1-$\sigma$ confidence level of each other ($2.02 \pm 0.28$ from \citet{REMY-RUYER14}). Unfortunately, the turnover point of broken power law derived in \citet{REMY-RUYER14} is at $\metal = 8.10 \pm 0.43$, and we do not have enough reliable DGR fitting results below that metallicity to compare with. It is hard to draw a conclusion whether a broken power law with turnover point around $\metal=8.0$ would fit our results better than a power law. The \citet{REMY-RUYER14} broken power law in high metallicity region is basically identical to the \citep{DRAINE11} power law.

\subsection{Comparison with MW depletion\label{sec: J09}}
\begin{figure}
\centering
\includegraphics[width=\columnwidth]{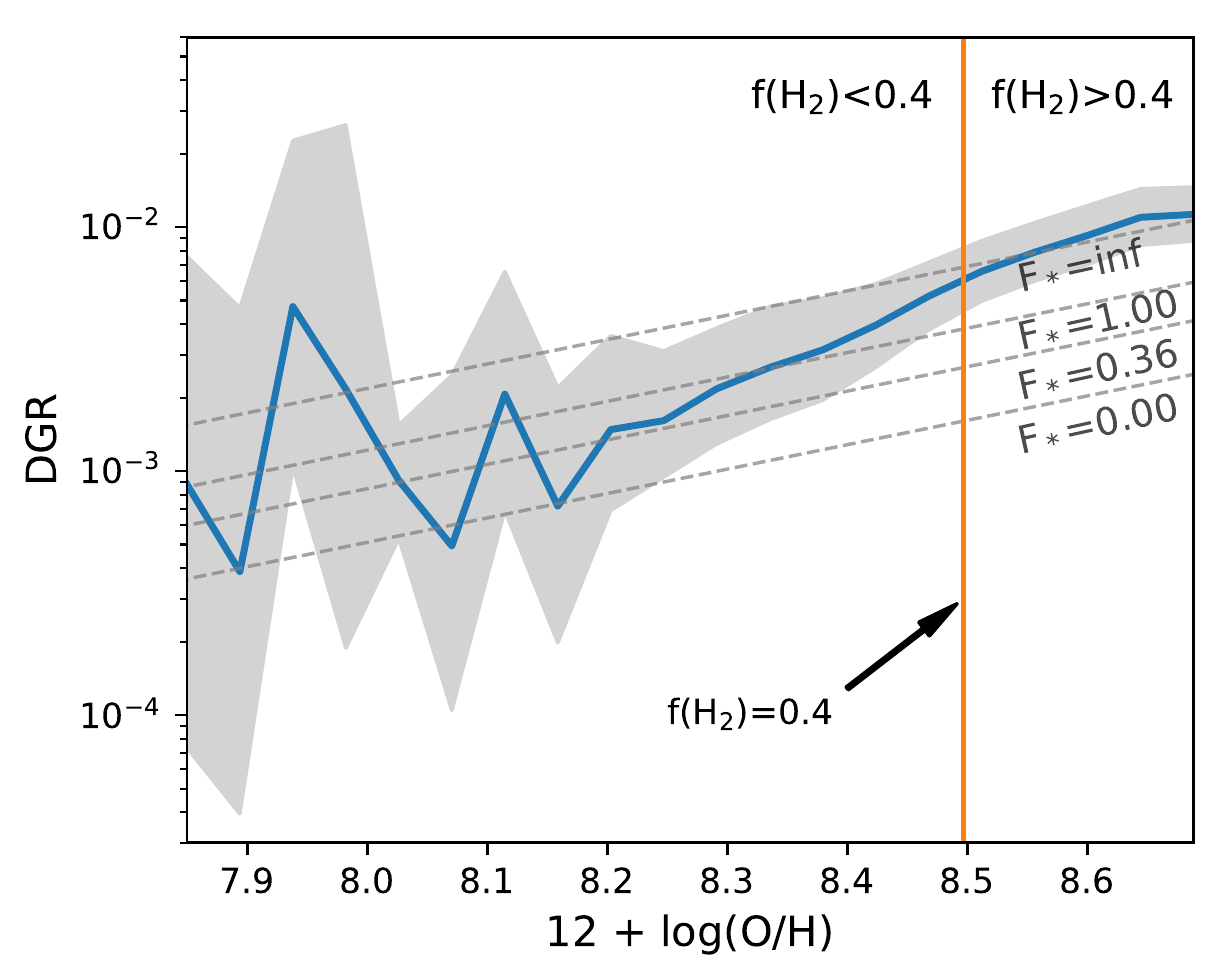}
\caption{The comparison of our results with the DTM corresponding to various MW $\rm F_\star$ values described in \citet{JENKINS09}. Most MW measurable regions have $0 \la {\rm F_\star} \la 1$. ${\rm F_\star}=0.36$ represents the average property of our $\kappa_{160}$ calibration, and ${\rm F_\star}=\rm inf$ means total depletion. The 40\% $\rm H_2$ location is marked because all \citet{JENKINS09} data points have $f_{\rm H_2} \la 0.4$.\label{fig: DTM_J09}}
\end{figure}
Studies of the depletion of heavy elements in the MW \citep{JENKINS09} also found a dependence of DTM on average ISM density and $\rm f_{H_2}$. In Figure \ref{fig: DTM_J09}, we display DTM corresponding to various MW $F_\star$ regions described in \citet{JENKINS09}. All of their original data points have $f_{\rm H_2} \la 0.4$ and $17.4 \la \logt (N_{\rm HI}) \la 21.8$. Regions with ${\rm F_\star}=1$ and ${\rm F_\star}=0$ are by definition the representative regions of high and low depletion in the diffuse ISM of the MW, respectively. Thus, the region between these two lines corresponds to a DTM similar to the MW range extending to lower metallicity. Most points with $\metal \leq 8.4$ fall inside this range. The high-latitude diffuse ISM in the MW used to calibrate our $\kappa_{160}$ has an ${\rm F_\star}$ of 0.36, thus it was selected for DGR calculation in calibrating our $\kappa_{160}$, see \secref{Sec: calibration}. The ${\rm F_\star}=\rm inf$ line means total depletion, which is physically the same as the DGR upper limit discussed in \secref{sec: max DGR}. All our DGR fitting results are within this limit. It is interesting to note that the point where the DGR trend falls below the maximum depletion is at the boundary of molecular gas dominant region and atomic gas dominant region (f$_{\rm H_2}\sim$0.4).

\subsection{Sensitivity of results to fitting methods}\label{sec:sensitivity}
It is worth noting that given the same dust emission SED, the fitting results are sensitive to methods and parameters in the fitting process. Thus, it is important to be clear and self-consistent about the choices we make for calibration and fitting, as demonstrated by \citet{GORDON14}. We also need to be careful when comparing cross-study results. Here, we use the process of $\kappa_{160}$ calibration with SE model, which gives $\kappa_{160}=10.48\pm1.48~\rm cm^2~g^{-1}$ with the SED of the MW diffuse ISM from \citet{GORDON14}, to illustrate the possible variations in results due to different choices. Note that we want to focus only on the methods, thus we use the MW diffuse ISM from \citet{GORDON14} in this section instead of ours described in \secref{Sec: calibration} to eliminate the simple offset.
\begin{itemize}
    \item By changing to different models, $\kappa_{160}$ can go up to 21.16 (PL model), which is a 100\% change. Thus, the choice of fitting model strongly affects fitting results.
    \item By making the fitting grid spacing coarser, from the original 0.002 spacing to a 0.1 spacing in $\logt\kappa_{160}$, the resulting $\kappa_{160}$ becomes 11.7, which is a 10\% change. This has a mild effect on fitting results, and is especially important when the grid spacing is larger than the adopted uncertainties.
    \item The matrix form and values of the covariance matrix can affect the fitting results. By changing the covariance matrix from ours to the one in \citet{GORDON14} and keeping all other factors the same, the resulting $\kappa_{160}$ goes to 17.9, which is a 70\% change. This also affects the results strongly.
    \item The covariance matrix can also change the fitting residuals. For example, \citet{GORDON14} assumes a flat uncertainty across the five bands and equal correlation, which results in similar residuals among the five bands. On the other hand, we assume different values and correlation between DIRBE and FIRAS bands, which results in better residuals in FIRAS bands and worse residual in the DIRBE band.
\end{itemize}

\section{Conclusions} \label{sec: conclusions}
We present dust SED fitting results from five MBB variants in M101 with kpc scale spatial resolution. We compare the resulting $\Sigma_d$ and $T_d$ with known physical limitations, and conclude the results from a simple, variable emissivity, modified blackbody model are not physically plausible. The other four models have results consistent with each other at $\metal \leq 8.5$, which demonstrates the robustness of modified blackbody model under many conditions. Among the four models, the one with a single temperature blackbody modified by a broken power-law emissivity has the highest fitting quality in residuals and $\tilde{\chi}^2$ distribution. Thus, the first order correction to the MBB, necessitated by our observed SEDs in M101, is to add flexibility in the emissivity spectral index at long wavelengths.

The resulting DTM, derived from our dust and gas surface densities and direct $\rm T_e$-based metallicities, is not constant with radius or metallicity in M101 from all five models. From the preferred BE model, a relation of $\rm DGR \propto Z^{1.7}$ is observed overall, and $\rm DGR \propto Z^{1.9}$ in region with $\metal \geq 8.2$. We try to explain this variable DTM by searching for correlations between tracers of formation and destruction mechanisms of dust and metallicity to the observed physical quantities. By comparing the correlation between DTM and physical quantities ($\rm f_{H_2}$, $\Sigma_\star$ and $\Sigma_{\rm SFR}$) after removing the major radial trend, we argue that the accretion of metals in ISM onto existing dust grains could be a cause of this variable DTM, while we do not see evidence for correlations with stellar or SNe related production and destruction.

It is also possible that the change in DTM is actually the enhancement of emissivity due to coagulation. In the center of M101, if we assume the \citet{DRAINE14} DTM and calculate the possible change in emissivity, the resulting $\kappa_{160}$ would be $\sim$19 to $41~\rm cm^2~g^{-1}$, which are 0.9 to 2.0 larger than the originally calibrated value of $16.52~\rm cm^2~g^{-1}$ in the high latitude diffuse ISM in the MW. This change is still within the range of previous observational and theoretical calculations. Both changes in DTM and in emissivity are possible according to our current knowledge.

When comparing with previous DTM studies, our DTM is 2 times larger than the \citet{DRAINE14} results in the same metallicity region, but our DTM are consistent with their DTM extrapolated to the region with similar $\rm f_{H_2}$. Comparing with \citet{REMY-RUYER14}, our DTM has a slope consistent with their power-law fitting slope. Unfortunately, we do not have enough low-metallicity data to compare with their broken-power law. When comparing with known depletion relations from the MW and the amount of available metals in the central 5 kpc of M101, our DTM suggests essentially all available heavy elements are in dust, which is consistent with ${\rm F_\star}={\rm inf}$ line from extrapolating the \citet{JENKINS09} calculations, and also larger than most of the previous studies. Our DTM results in the lower metallicity region would fall between ${\rm F_\star}=1$ and ${\rm F_\star}=0$ in the MW. This suggests that even in the lowest metallicity regime of our study, we have not yet probed conditions where the dust life cycle differs in major ways from that in the Milky Way.

During the fitting process, we found that the fitting results from the likelihood calculated with a multi-dimensional Gaussian distribution and a complete covariance matrix are sensitive to the choice of model and covariance matrix. Therefore, it is important to be self-consistent between calibration and fitting processes. It is also important to note the covariance matrix adopted when comparing fitting results across studies because the fitting results could change by 70\% with different covariance matrices.

\acknowledgments
We thank the referee for useful comments that helped to improve the quality of the manuscript. We gratefully acknowledge the hard work of the KINGFISH, THINGS, HERACLES, LVL, and CHAOS teams and thank them for making their data publicly available. We acknowledge the usage of the HyperLeda database (http://leda.univ-lyon1.fr). IC thanks K. Gordon for helpful conversations regarding calibration and fitting. IC thanks Y.-C. Chen for helpful conversations. The work of KS, IC, AKL, DU and JC is supported by National Science Foundation grant No. 1615728 and NASA ADAP grants NNX16AF48G and NNX17AF39G. The work of AKL and DU is partially supported by the National Science Foundation under Grants No. 1615105, 1615109, and 1653300.

This work uses observations made with \textit{Herschel}. \textit{Herschel} is an ESA space observatory with science instruments provided by European-led Principal Investigator consortia and with important participation from NASA. 
PACS has been developed by a consortium of institutes led by MPE (Germany) and including UVIE (Austria); KU Leuven, CSL, IMEC (Belgium); CEA, LAM (France); MPIA (Germany); INAF-IFSI/OAA/OAP/OAT, LENS, SISSA (Italy); IAC (Spain). This development has been supported by the funding agencies BMVIT (Austria), ESA-PRODEX (Belgium), CEA/CNES (France), DLR (Germany), ASI/INAF (Italy), and CICYT/MCYT (Spain). 
SPIRE has been developed by a consortium of institutes led by Cardiff University (UK) and including Univ. Lethbridge (Canada); NAOC (China); CEA, LAM (France); IFSI, Univ. Padua (Italy); IAC (Spain); Stockholm Observatory (Sweden); Imperial College London, RAL, UCL-MSSL, UKATC, Univ. Sussex (UK); and Caltech, JPL, NHSC, Univ. Colorado (USA). This development has been supported by national funding agencies: CSA (Canada); NAOC (China); CEA, CNES, CNRS (France); ASI (Italy); MCINN (Spain); SNSB (Sweden); STFC, UKSA (UK); and NASA (USA).

This work uses observations based on National Radio Astronomy Observatory (NRAO) Karl G. Jansky Very Large Array. The NRAO is a facility of the National Science Foundation operated under cooperative agreement by Associated Universities, Inc.
This work uses observations based on HERA on IRAM 30-m telescope. IRAM is supported by CNRS/INSU (France), the MPG (Ger-
many) and the IGN (Spain). 
This work uses observations made with the \textit{Spitzer} Space Telescope, which is operated by the Jet Propulsion Laboratory, California Institute of Technology under a contract with NASA. 
This research has made use of NASA's Astrophysics Data System. This research has made use of the NASA/IPAC Extragalactic Database (NED) which is operated by the Jet Propulsion Laboratory, California Institute of Technology, under contract with the National Aeronautics and Space Administration.

\vspace{5mm}
\facilities{Herschel(PACS and SPIRE), VLA, GALEX, IRAM(HERA), Spitzer(MIPS and IRAC), LBT(MODS)}
\software{astropy \citep{ASTROPY13},
          matplotlib \citep{HUNTER07},
          numpy \& scipy \citep{VANDERWALT11},
          pandas \citep{MCKINNEY10},
          voronoi\_2d\_binning \citep{CAPPELLARI03},
          corner \citep{Foreman-Mackey16}, 
          Scanamorphos \citep[v16.9 and v17.0;][]{ROUSSEL13}, 
          HIPE \citep[vspire-8.0.3287;][]{OTT10}}

\appendix
\section{Full fitting results\label{app: fitting}}
\begin{figure}[htb!]
\gridline{\fig{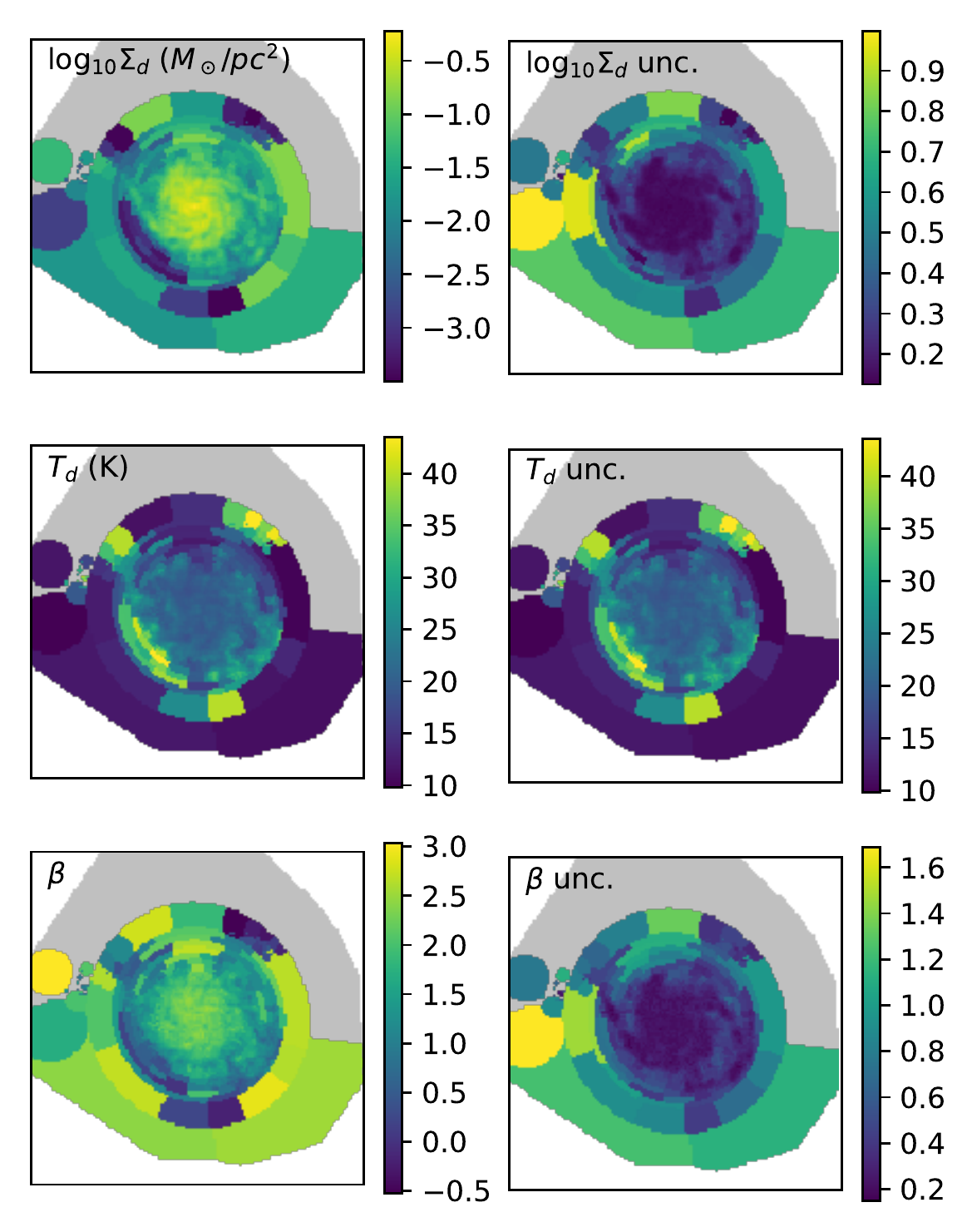}{0.45\textwidth}{(a) SE model}
          \fig{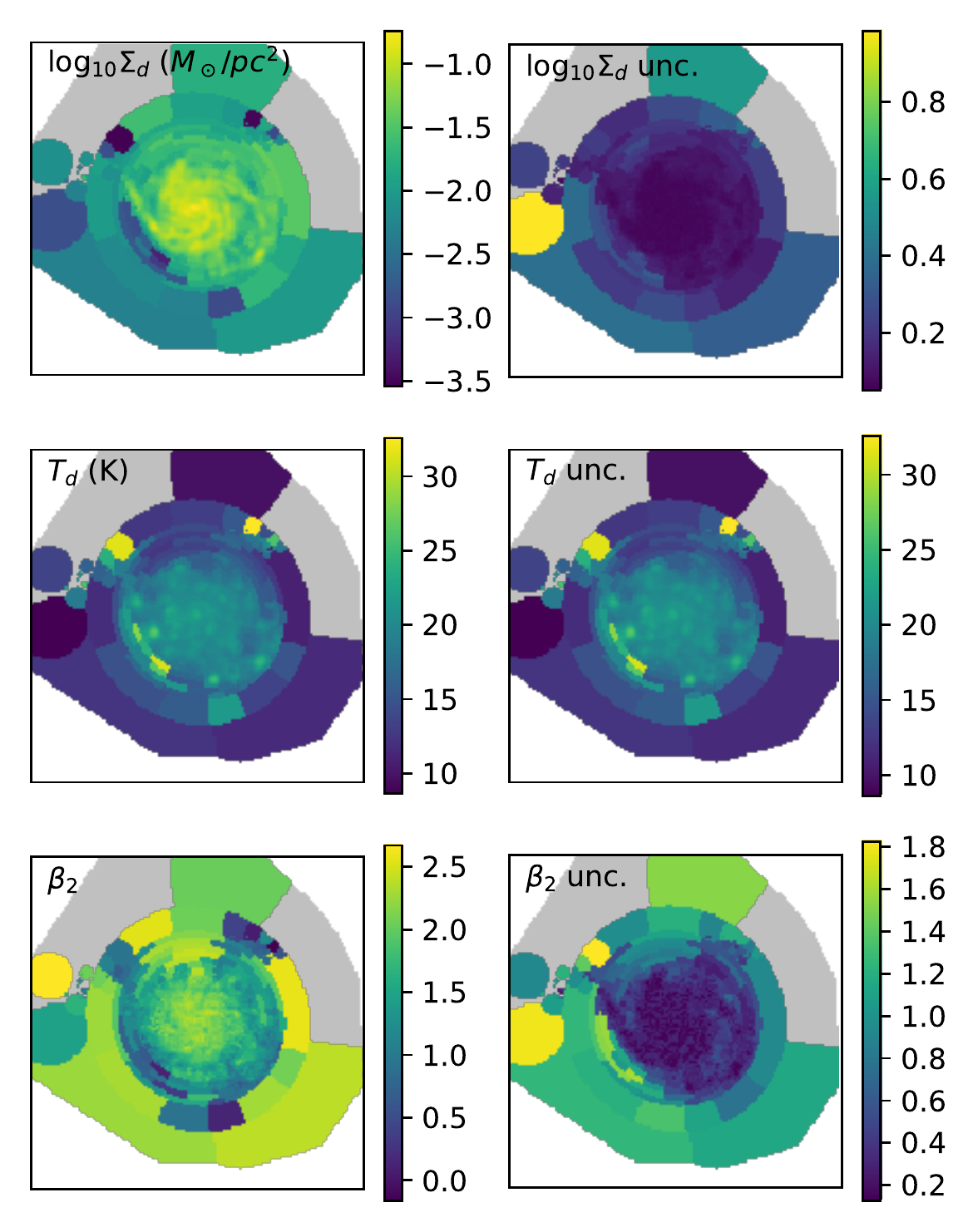}{0.45\textwidth}{(b) BE model}
          }
\gridline{\fig{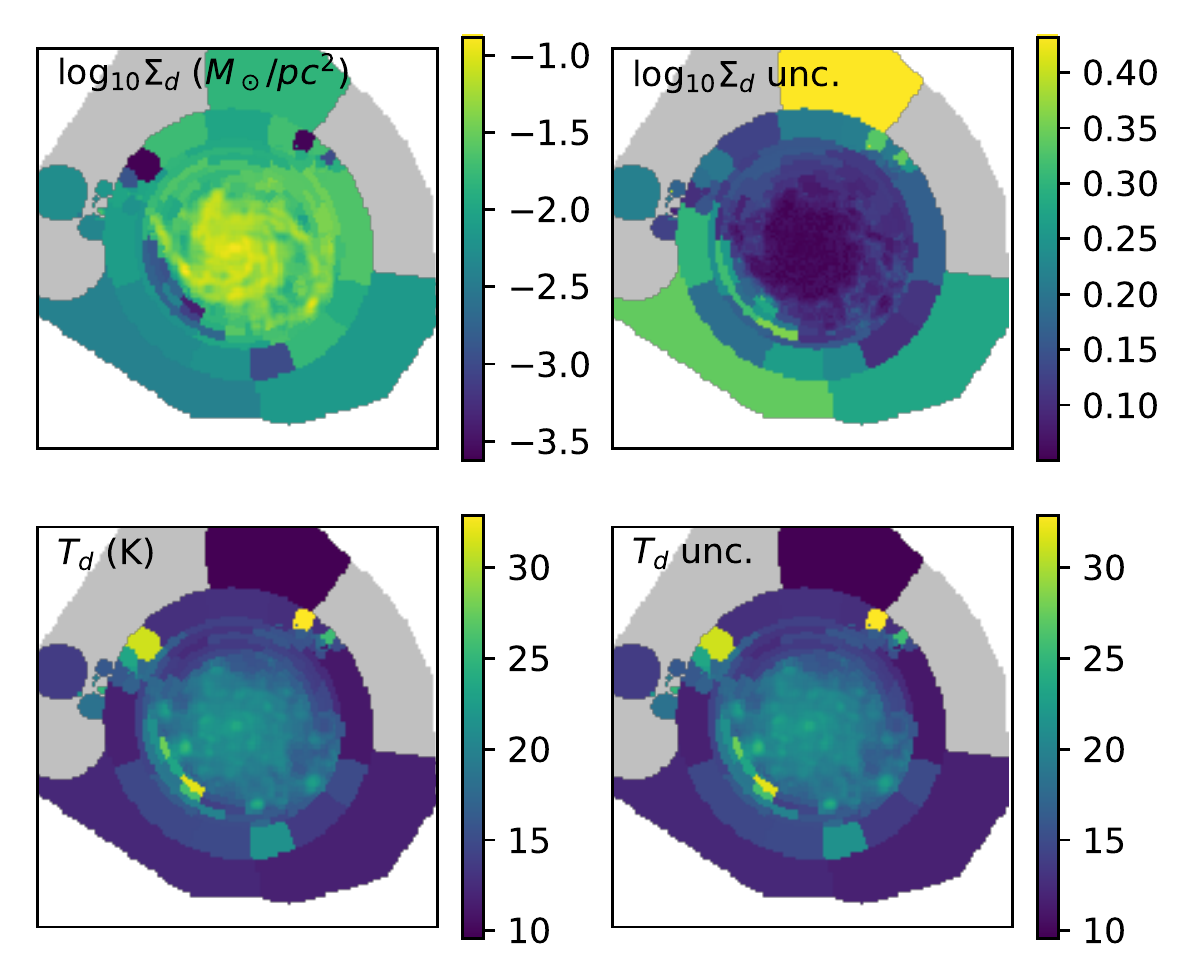}{0.45\textwidth}{(c) FB model}
          }
\caption{Fitting results from (a) SE model. (b) BE model. (c) FB model. The left panels show the maps of the parameters, and the right panels show the corresponding fitting uncertainties. The gray region is the poorest fit region, where the uncertainties in $\Sigma_d$ are larger than 1 dex.\label{fig: results_SEBEFB}}
\end{figure}

\begin{figure}[htb!]
\gridline{\fig{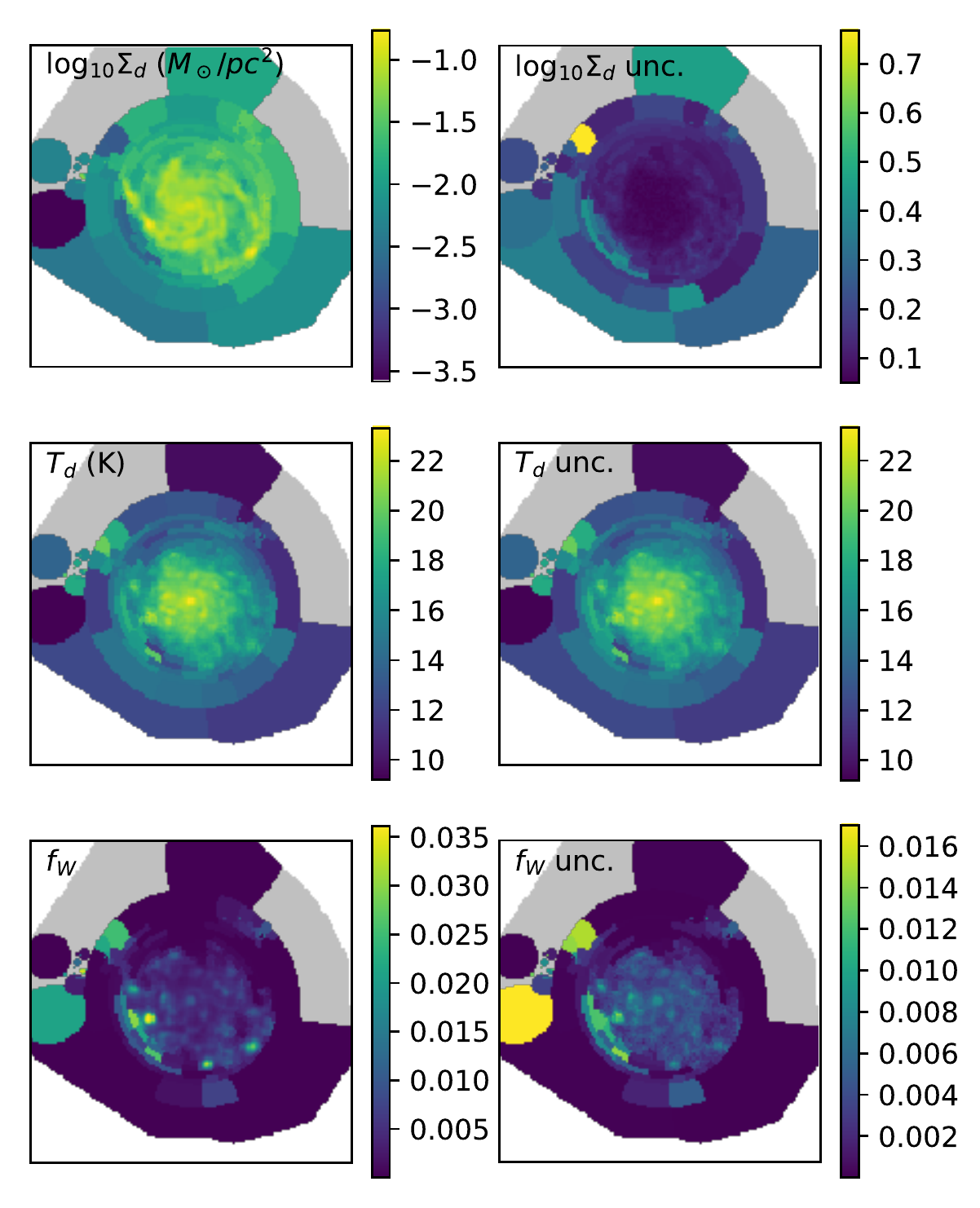}{0.45\textwidth}{(a) WD model}
          \fig{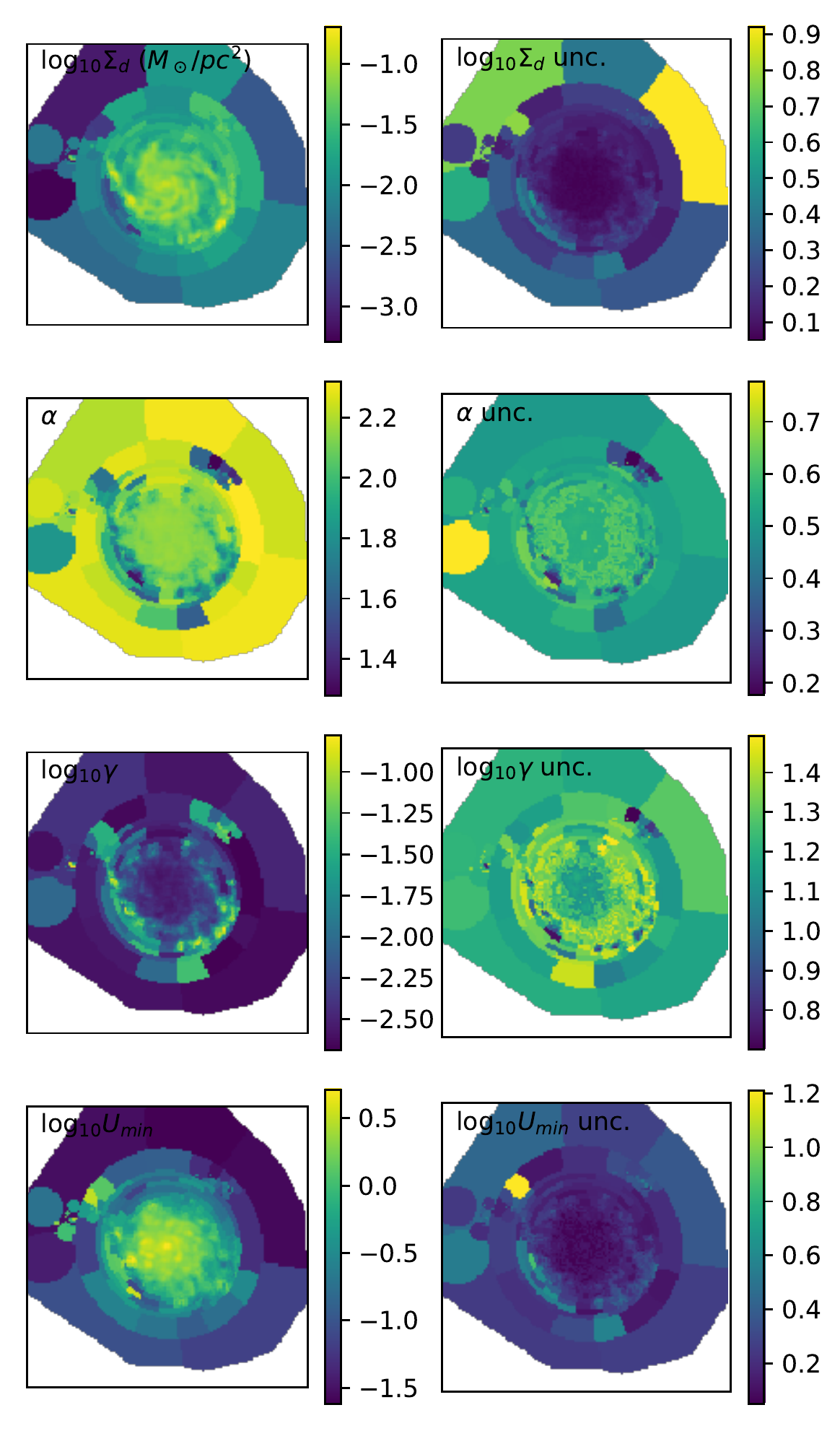}{0.45\textwidth}{(b) PL model}
          }
\caption{Fitting results from (a) WD model. (b) PL model. The left panels show the maps of the parameters, and the right panels show the corresponding fitting uncertainties. The gray region is the poorest fit region, where the uncertainties in $\Sigma_d$ are larger than 1 dex.\label{fig: results_WDPL}}
\end{figure}

\begin{figure}[htb!]
\centering
\includegraphics[width=0.7\columnwidth]{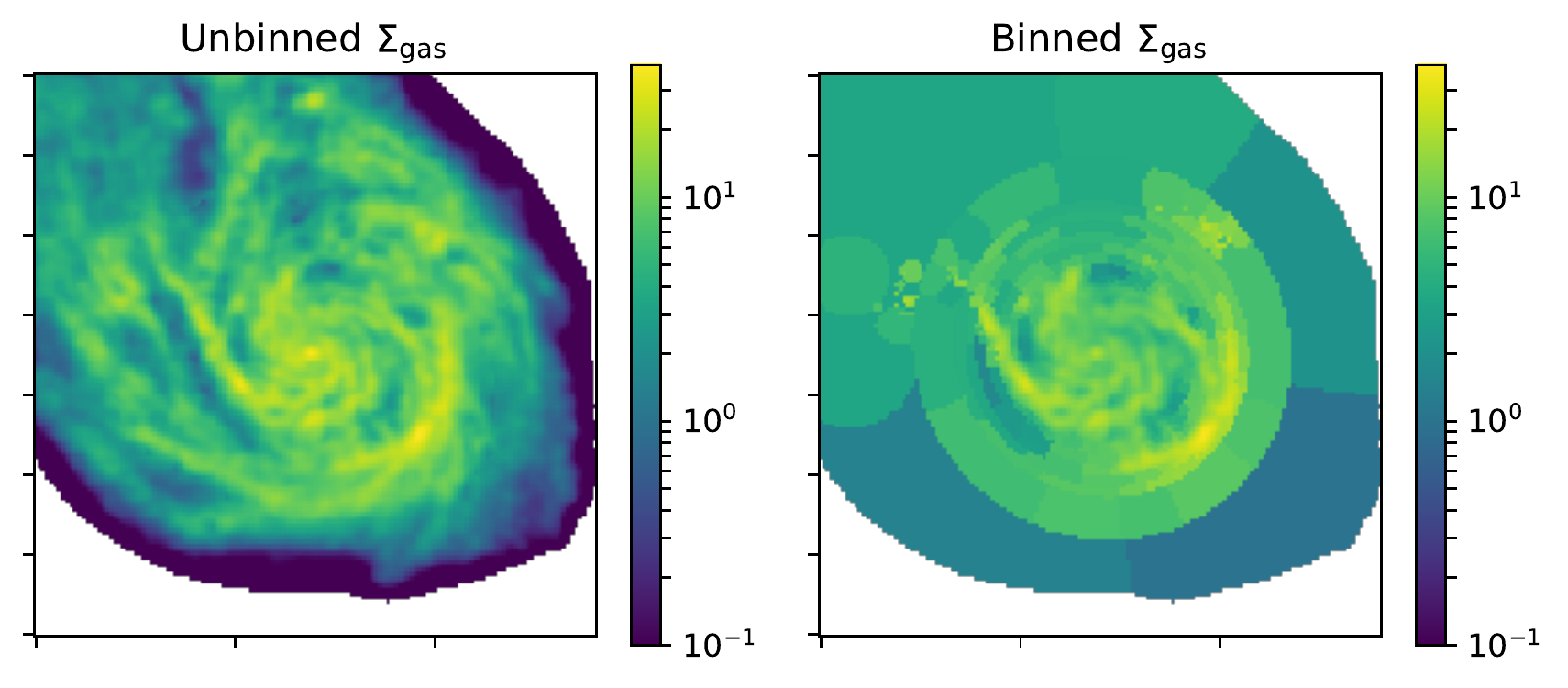}
\caption{The spatial distribution of $\Sigma_{\rm gas}$ ($M_\sun~\rm pc^{-2}$). Left: The distribution at unbinned SPIRE500 resolution. Right: The binned distribution.\label{fig: binned_unbinned_gas}}
\end{figure}

The full fitting results from SE, FB, WD, and PL are shown in Figure \ref{fig: results_SEBEFB}-\ref{fig: results_WDPL}. The white areas are the background regions, where the SED are not fitted. The gray area is the poorest fit region, where the uncertainty in $\Sigma_d$ is larger than 1 dex. The fitting uncertainties are shown along with the fitted values. The discontinuities in the binned data result from the change in ISM surface density, which is demonstrated in Figure \ref{fig: binned_unbinned_gas} with a comparison between binned and unbinned $\Sigma_{\rm gas}$ maps.

\section{Correlation between fitting parameters\label{app: corner}}
\begin{figure*}[htb!]
\gridline{\leftfig{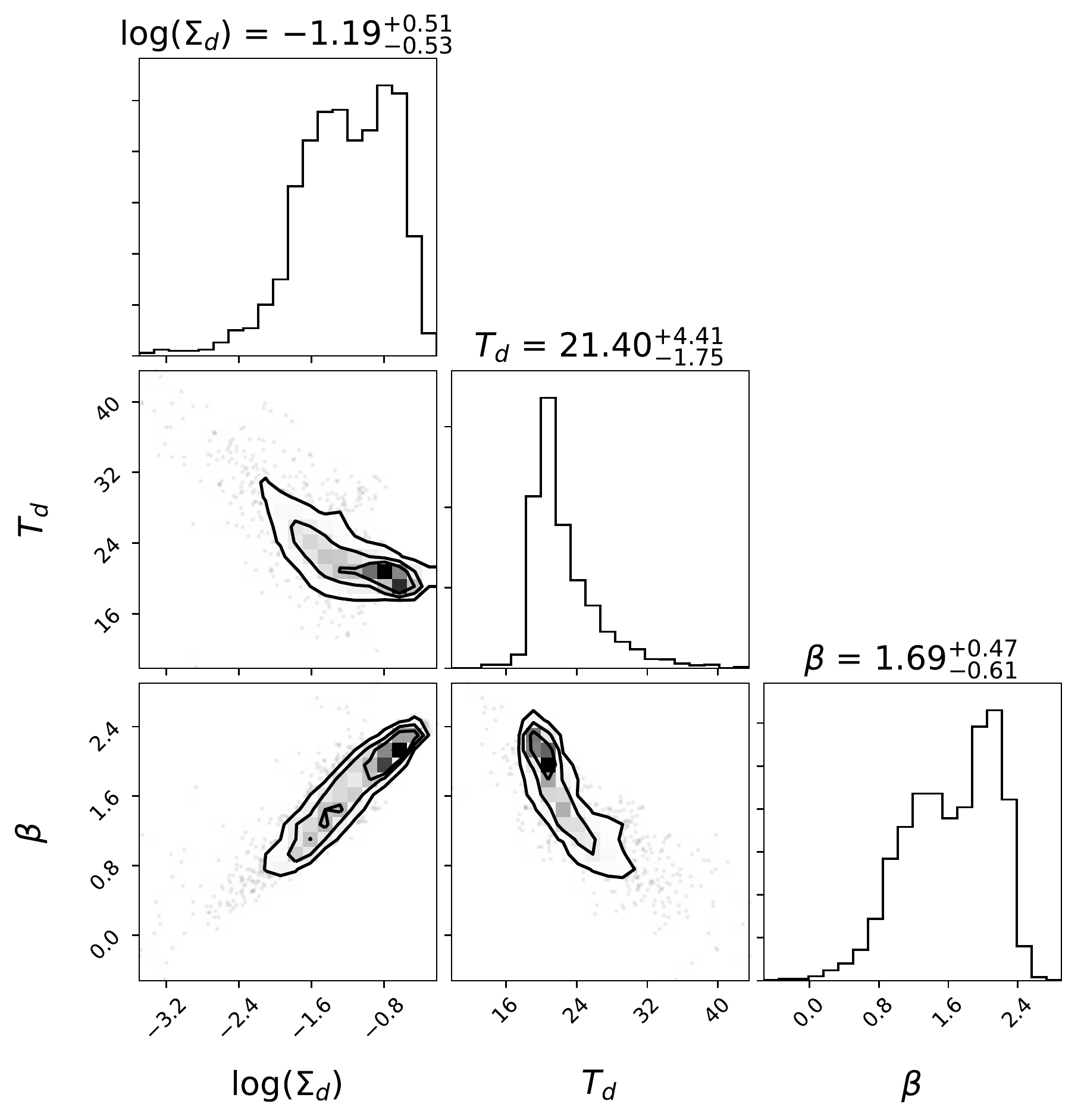}{0.45\textwidth}{(a) SE model}
          \fig{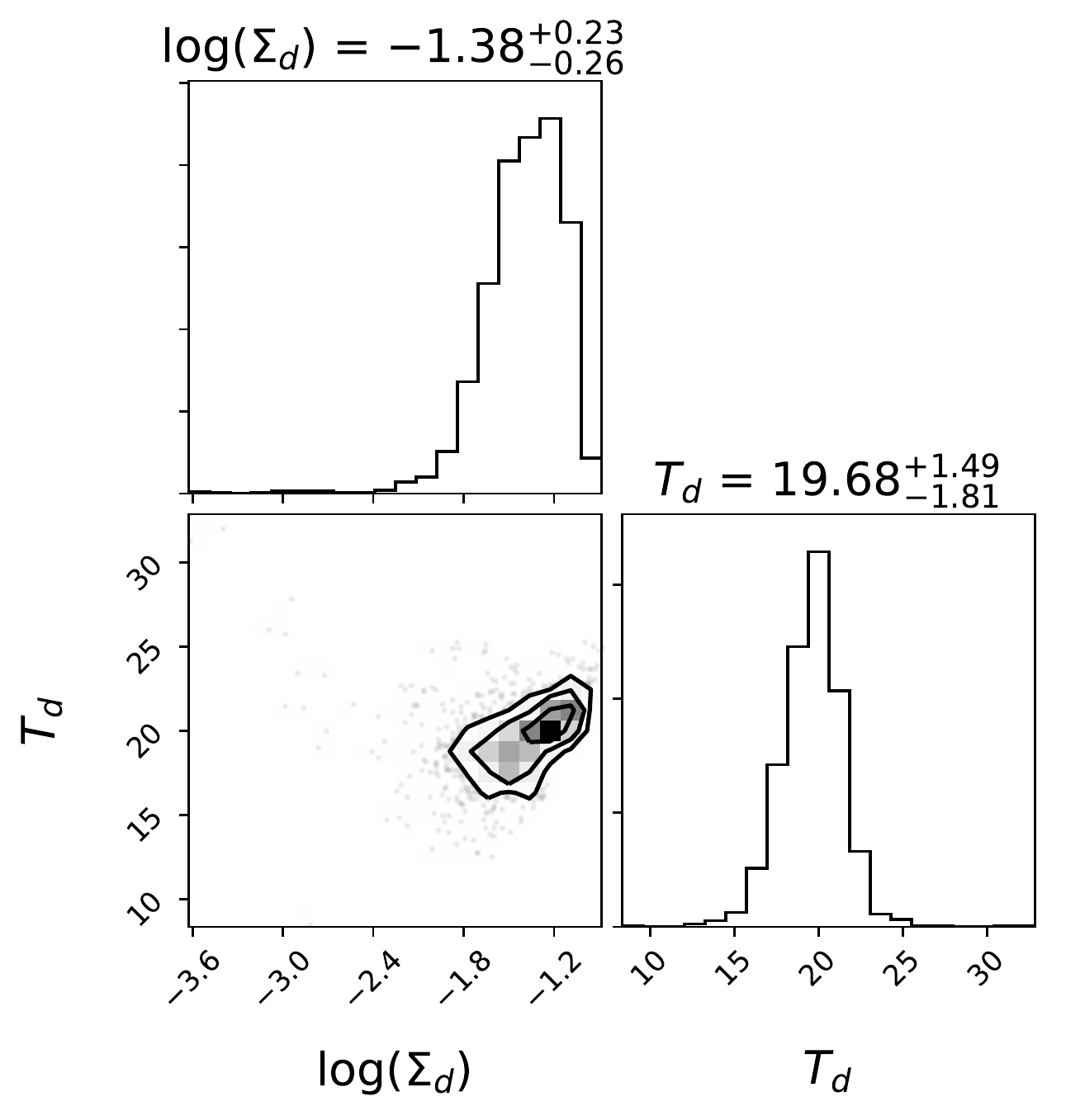}{0.3\textwidth}{(b) FB model}
          }
\gridline{\leftfig{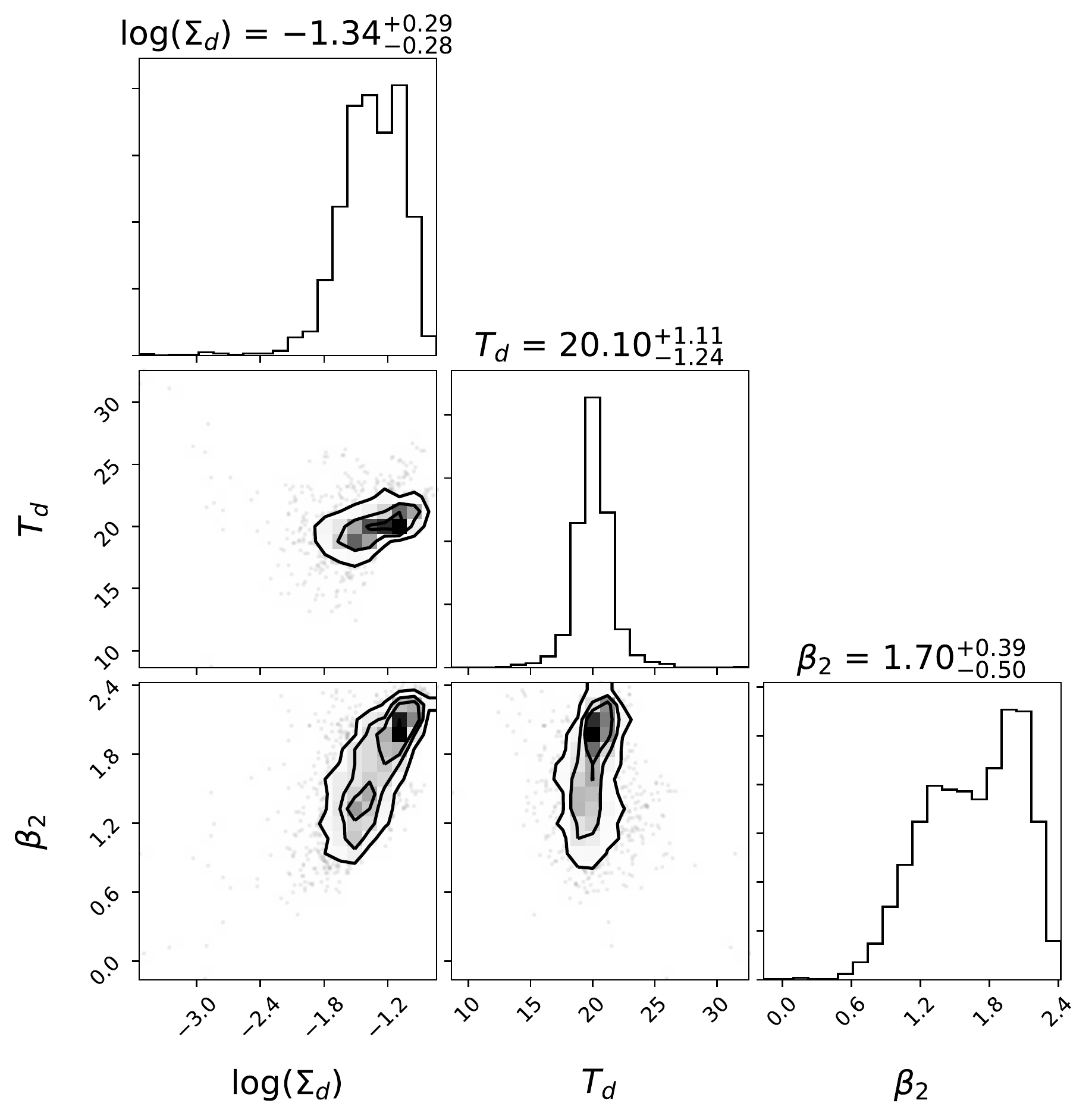}{0.45\textwidth}{(c) BE model}
          \rightfig{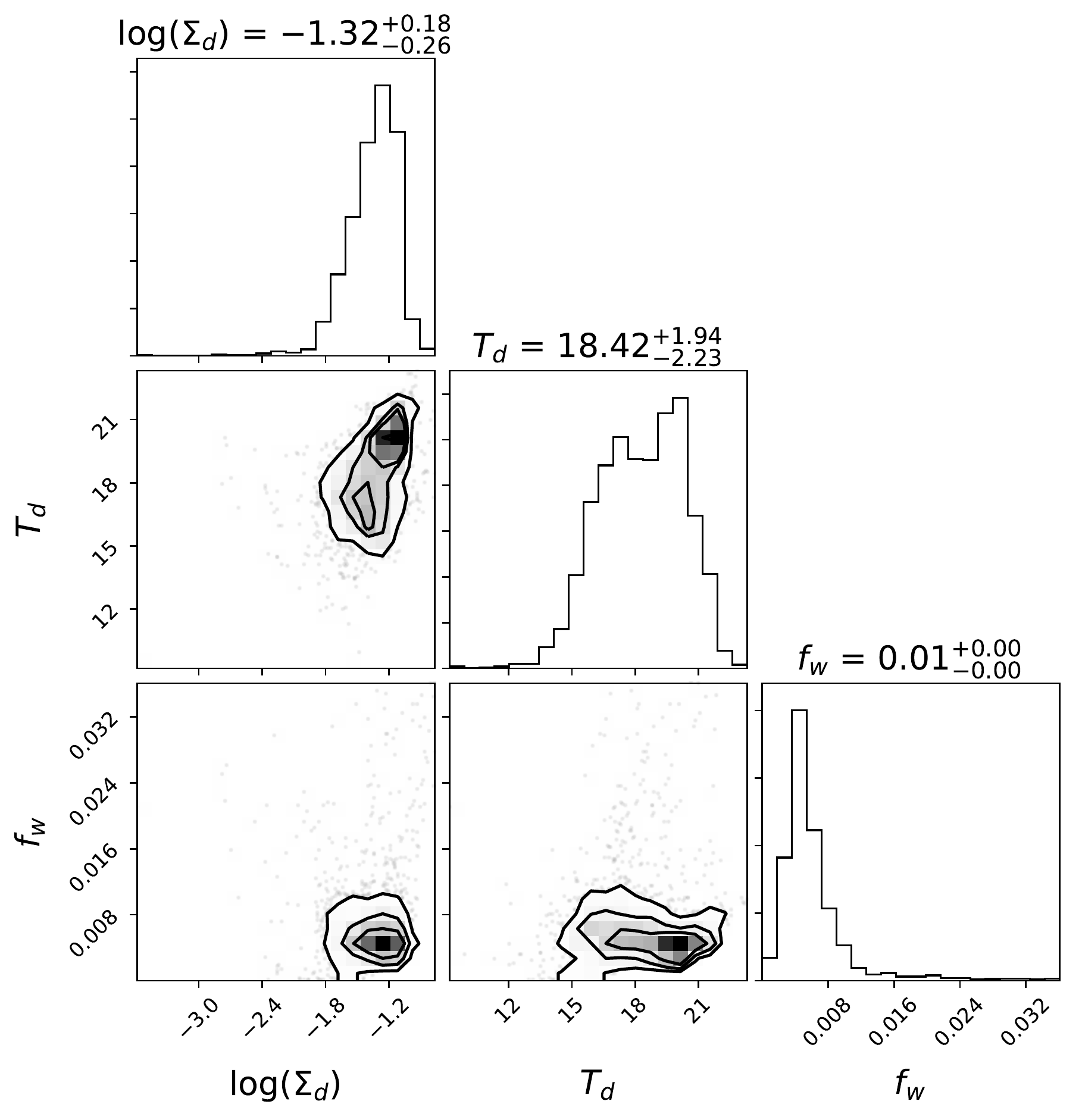}{0.45\textwidth}{(d) WD model}
          }
\caption{Correlation between parameters from fitting results. The histogram and 2-dimensional histograms show the distribution of expectation values of each parameter from each of the binned region. The values in the titles are the median and 16-84 percentile. (a) SE mode. (b) FB model. (c) BE model. (d) WD model.\label{fig: corner_SEFBBEWD}}
\end{figure*}

\begin{figure*}[htb!]
\gridline{\fig{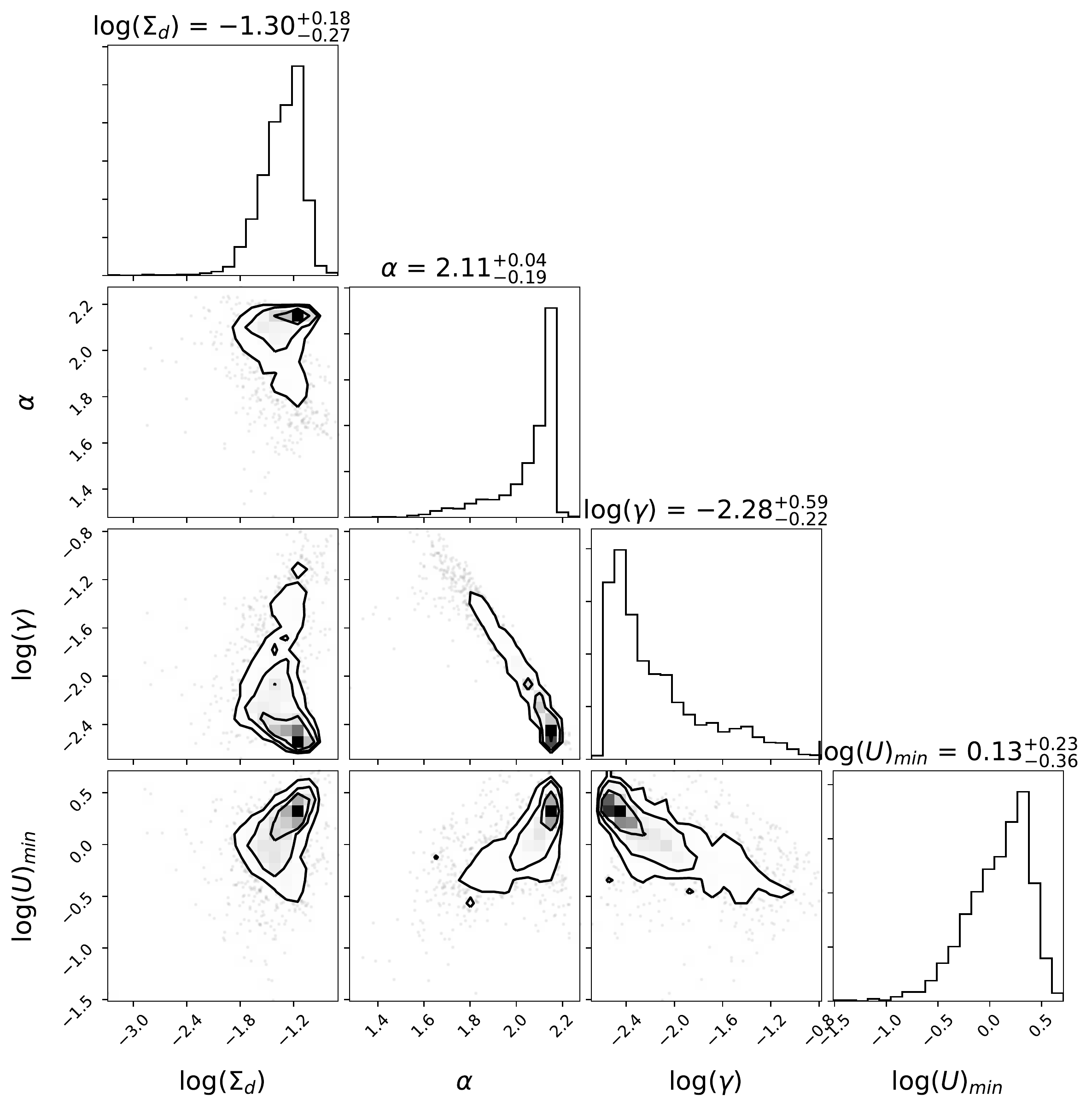}{0.65\textwidth}{}
          }
\caption{Same as Figure \ref{fig: corner_SEFBBEWD}, but for PL model.\label{fig: corner_PL}}
\end{figure*}

We plot the correlation between parameters in Figure \ref{fig: corner_SEFBBEWD}-\ref{fig: corner_PL}. The histogram and 2-dimensional histograms show the distribution of expectation value for each parameter from each of the binned region. The values in the titles are the median and 16-84 percentile.

\bibliographystyle{aasjournal}
\bibliography{idchiang}

%\listofchanges
\end{document}